# Valley-controlled many-body exciton interactions in monolayer $WSe_2$ phototransistors

*Daniel Vaquero[1]*, Cédric A. Cordero-Silis[1], Daniel Erkensten[2], Roberto Rosati[2], Martijn H. Takens[1], Kenji Watanabe[3], Takashi Taniguchi[4], Ermin Malic[2], Marcos H. D. Guimarães[1]*.*

[1]Zernike Institute for Advanced Materials, University of Groningen, 9747 AG Groningen, The Netherlands

[2]Department of Physics, Philipps-Universität Marburg, 35037 Marburg, Germany.

[3]Research Center for Electronic and Optical Materials, National Institute for Materials Science, 1-1 Namiki, Tsukuba 305-0044, Japan

[4]Research Center for Materials Nanoarchitectonics, National Institute for Materials Science, 1-1 Namiki, Tsukuba 305-0044, Japan

ABSTRACT. Many-body exciton interactions shape the optoelectronic response of atomically-thin transition metal dichalcogenides, yet optical control of these interactions remains largely unexplored. To date, modulation of exciton-exciton interactions has primarily relied on electrical gating or van der Waals engineering. Here, we demonstrate all-optical control of many-body exciton interactions in monolayer $WSe_2$ via valley-selective excitation using polarization-resolved pulsed-laser photocurrent spectroscopy. Circular excitation selectively populates excitons in a single valley, whereas linear excitation populates both valleys, inducing a valley-dependent nonlinear photoresponse. We observe helicity-dependent exciton renormalization, alongside a two-fold enhancement of sublinear photocurrent scaling under circular excitation, reflecting single-valley population of interacting excitons. A microscopic model incorporating intervalley-exchange and exciton-exciton annihilation mediated by dark and bright exciton populations reproduces the nonlinear valley-selective response. These results establish the valley degree of freedom as an all-optical control parameter for tuning many-body excitonic effects and, exploring correlated exciton states and valleytronic applications in two-dimensional semiconductors.

## MAIN TEXT

Many-body interactions in quantum-confined systems strongly influence their optoelectronic properties by reshaping the light-matter interaction at the nanoscale. A prominent example of this coupling in condensed matter physics is the formation of excitons, bound quasiparticles formed by electron-hole pairs through Coulomb interaction following light excitation across the bandgap of a semiconductor. Their population can be precisely tuned through optical pumping, providing a versatile platform for investigating many-body dynamics in low-dimensional systems.

Monolayer (1L) transition metal dichalcogenides (TMDs) have emerged as a paradigm for exploring rich many-body excitonic phenomena due to their direct band gap, two-dimensional confinement and strong Coulomb interactions[1]. These materials exhibit enhanced light-matter interaction, hosting tightly bound excitons with binding energies exceeding 100 meV, and lifetimes reaching hundreds of picoseconds even at room temperature[2,3]. Such properties enable access to high excitonic densities – up to $10^{13}$-$10^{14}$ $cm^{-2}$ – without reaching the Mott excitonic transition, allowing to probe many-body excitonic phenomena in the absence of free electron-hole plasma[4,5]. At high exciton densities, interactions between excitons become significant, giving rise to nonlinear processes such as exciton–exciton annihilation[6], exciton–exciton scattering[7], phase space filling, formation of spatial rings (halos)[8,9] and multiexcitonic states[10], making 1L-TMDs a unique platform to explore exciton interaction-driven physics.

Excitonic systems are strongly influenced by factors such as the dielectric environment[11] and exciton species. In van der Waals materials, stacking, twisting, and moiré trapping[12,13] reshape exciton dimensionality, screening, and binding energies through interlayer hybridization[14,15]. Although these strategies tune many-body interactions, they require modifying the semiconductor itself rather than enabling dynamic control over it. In 1L-TMDs, broken inversion symmetry and strong spin-orbit coupling impose valley-selective optical selection rules[16]. Circularly polarized light of opposite helicities promotes selectively excitons in the two inequivalent K and K′ valleys[17] providing a unique all-optical knob to control exciton interactions through this degree-of-freedom[18].

Here, we demonstrate the optical valleytronic control of many-body exciton interactions by selectively exciting carriers at the K and K' valleys of the Brillouin zone in 1L-$WSe_2$. Using polarization-resolved power-dependent photocurrent spectroscopy, we reveal pronounced sublinear photocurrent scaling under increasing laser fluence, a hallmark of many-body exciton interactions, and show that its strength depends on the light polarization and observable for temperatures up to 100 K. Using a microscopic model, which accounts for a temperature-dependent valley-scattering and exciton–exciton annihilation, we explain the temperature dependence on the valley-dependent exciton interactions. Previous helicity-dependent photocurrent studies in 1L-TMDs have mainly focused on circular photogalvanic effects or photon drag effects[19,20]. In contrast, this optical control over many-body effects opens the possibility for harnessing these phenomena to realize optically-controllable excitonic and valleytronic devices operating in regimes of high exciton density, where strong exciton–exciton interactions give rise to nonlinearities, collective phases such as excitonic condensates[21], and coherent light-matter states like polariton fluids[22].

**Many-body fingerprints in exciton-driven photoresponse**

Excitons dominate the photoresponse of 1L-TMDs based devices due to their large binding energies. Figure 1a displays the schematic of the graphite/hexagonal boron nitride (h-BN)/1L-$WSe_2$ phototransistor (see section S2) illuminated by linear (black) and circular (red) polarized light. We illuminate the device using ultrafast (100-fs) laser pulses that photoexcite excitons which are dissociated under a bias voltage between the electrodes generating photocurrent[23,24] (see Methods and section S3). The device presents an *n*-type transistor behaviour showing an increase in the conductivity as we increase the $V_g$ over the threshold voltage ($V_{th}$ = 5.3 V, see section S4). All the photocurrent measurements are performed at $V_{ds}$ = 5 V and $V_g$ = 0 V. In this regime, the photoresponse is governed primarily by photoconductive processes, and the photocurrent shows a linear dependence on illumination power[25,26].

The coupled spin–valley band structure of 1L-TMDs enables direct optical control of valley populations. Circularly polarized light ($\sigma^{\pm}$) excites carriers selectively in one of the inequivalent valleys (K or K′), whereas linearly polarized ($L$) light excites both valleys equally, since it can be decomposed into equal components of opposite helicities: $L = \frac{1}{\sqrt{2}}(\sigma^{+} + \sigma^{-})$. Figures 1b,c sketch that linear excitation creates equal exciton populations in K (spin up) and K′ (spin down), while circular excitation confines carriers to a single valley.

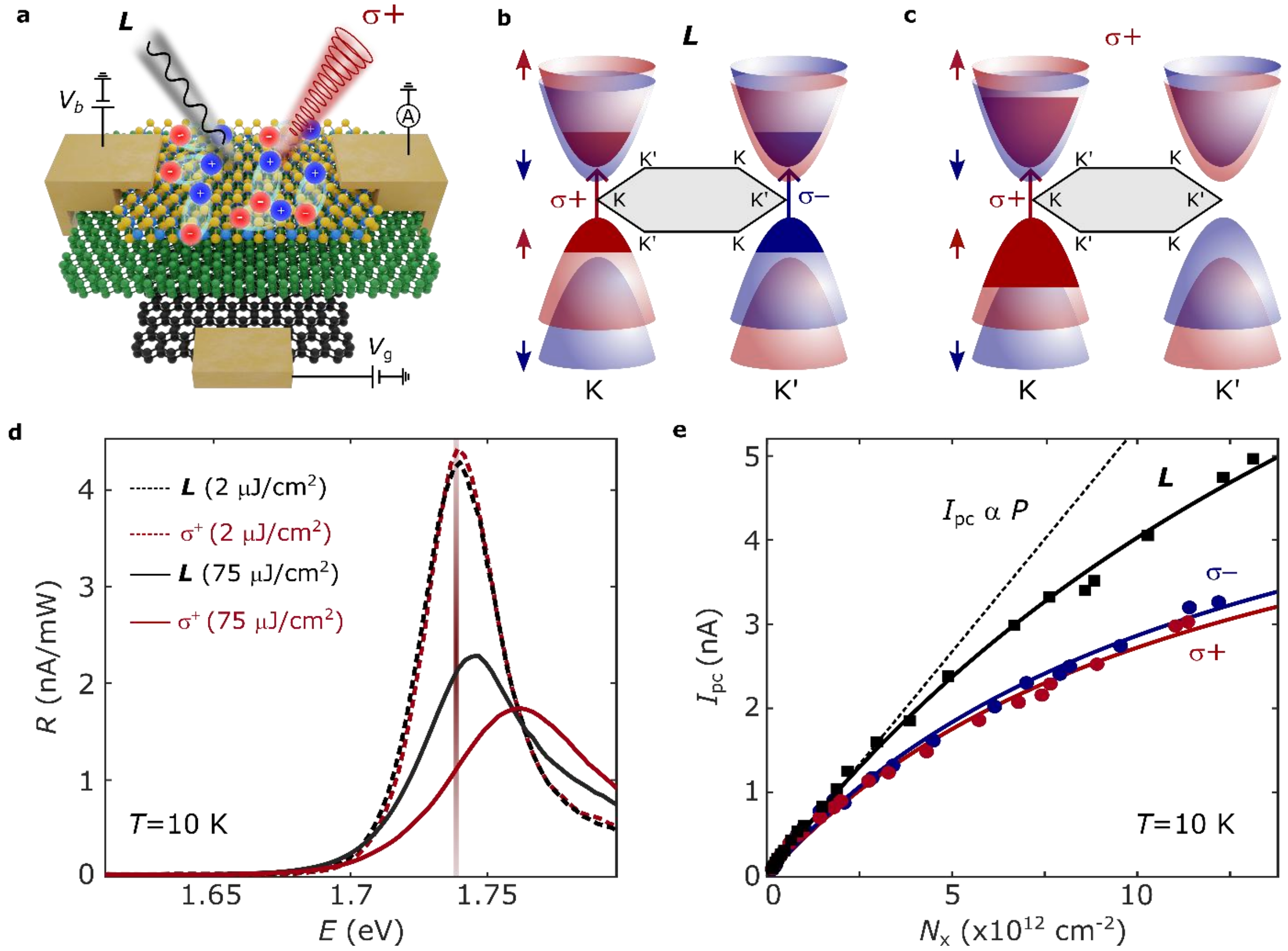

**Figure 1.** Valley controlled many-body effects in 1L-$WSe_2$ photodetector. (a**)** Schematic representation of photoexcited excitons in a graphite/hBN/1L-$WSe_2$ heterostructure contacted by Pt/Au electrodes. The graphite and h-BN layers are shown in black and green, respectively, while W and Se atoms are depicted in blue and yellow. A gate voltage $V_g$ is applied through the graphite back gate, and a drain-source bias $V_{ds}$ is applied between Pt/Au contacts used to measure the photocurrent. Band diagram sketch of 1L-$WSe_2$ with the exciton density in the different valleys for linear (b) and circular (c) incident light polarizations. (d) Responsivity spectra of the under linear ($L$ – black lines) and right-handed circular ($\sigma^+$ – red lines) excitation measured at low fluence (2 μJ/cm$^2$) and high fluence (75 μJ/cm$^2$) depicted with dashed and solid lines, respectively. (e) Photocurrent generated as a function of the exciton density ($N_X$) for linear and circular polarized light. The solid lines show the fitting of the experimental data using Equation 1.

For parabolic bands, the density-of-states is $g(E) = \frac{g_s g_v \mu}{2\pi\hbar^2}$, where $g_s$ and $g_v$ are the spin and valley degeneracy, and $\mu$ is the exciton reduced effective mass. Although the total degeneracy of available states remains unchanged, circular excitation concentrates excitons within a single valley, doubling the exciton density per valley compared with linear excitation at the same fluence. This characteristic spin-valley locked band structure significantly influences the optical response of 1L-TMDs at exciton densities ranging from $10^{11}$ to $10^{13}$ $cm^{-2}$, where many-body excitonic effects become dominant[7,27–29].

To characterize the valley-dependent exciton many-body effects, we measure the photocurrent spectra at different excitonic density regimes from 0.1 to $12\times10^{12}$ $cm^{-2}$, that correspond to pump fluences ranging between 0.5 – 75 μJ/$cm^2$, below the excitonic Mott transition estimated for 1L-$WSe_2$[30]. Figure 1d shows the responsivity spectra of the 1L-$WSe_2$ device acquired under linear ($L$ – black) and right-handed ($\sigma^+$ – red) polarized light at low (2 μJ/$cm^2$ – dashed), and high (75 μJ/$cm^2$ – solid) fluences. The A-exciton strongly dominates the photocurrent spectra at a photon energy of $E$ = 1.74 eV. At low fluences, both $L$ and $\sigma^+$, polarized spectra are nearly identical, however, at high fluences differ significantly. The A-exciton remains dominant but the overall lineshape undergoes substantial modifications, characterized by robust optical signatures of many-body interactions: reduced device responsivity, blueshift of the exciton resonance, and exciton linewidth broadening. These changes in spectral features reflect and confirm the role of many-body exciton interactions at elevated excitonic densities and drastically vary with the valley-dependent excitation. Circularly-polarized pumping –which concentrates excitons in a single valley– amplifies the interaction effects, leading to higher nonlinearities. In contrast, linearly-polarized excitation distributes the carriers between valleys and mitigates and their interactions.

Consistently, at high fluences, all three many-body signatures are more pronounced under circular polarized light excitation.

In a simple phase-space filling picture, concentrating excitons in a single valley enhances many-body interactions compared with distributing them across two valleys, leading to the stronger responsivity reduction under circular excitation (Fig. 1d). As the exciton density increases, exciton-exciton interactions become dominant, giving rise to a markedly sublinear photocurrent response. Figure 1e presents the value of the photocurrent power dependence measured in resonance with exciton A for $L$ (black squares), $\sigma^+$ (red dots), and $\sigma^-$ (blue dots) polarizations. For both, the photocurrent exhibits pronounced sublinear dependences on the excitation power that can be described by a standard saturation model[31]:

$$I_{\mathrm{pc}} = \frac{I_0}{1 + \frac{N_x^s}{N_x}}, \qquad (1)$$

where $I_0$ denotes the saturation value of the photocurrent and $N_x^s$ represents the exciton density at which $I_{\mathrm{pc}}$ reaches half of $I_0$. From the fittings in Fig 1e we obtain saturation currents $I_{0,\mathrm{L}} = 13.35 \pm 0.98$ nA, $I_{0,\sigma^+} = 6.08 \pm 0.43$ nA, and $I_{0,\sigma^-} = 6.52 \pm 0.42$ nA, corresponding to ratios $\frac{I_{0,\mathrm{L}}}{I_{0,\sigma^+}} = 2.19 \pm 0.22$ and $\frac{I_{0,\mathrm{L}}}{I_{0,\sigma^-}} = 2.05 \pm 0.20$. The saturation densities obtained are $N_{x,\mathrm{L}}^{\mathrm{s}} = (2.33 \pm 0.23) \times 10^{13}\ \mathrm{cm}^{-2}$, $N_{x,\sigma^+}^{\mathrm{s}} = (1.21 \pm 0.14) \times 10^{13}\ \mathrm{cm}^{-2}$ and $N_{x,\sigma^-}^{\mathrm{s}} = (1.24 \pm 0.13) \times 10^{13}\ \mathrm{cm}^{-2}$, with ratios $\frac{N_{x,\mathrm{L}}^{\mathrm{s}}}{N_{x,\sigma^+}^{\mathrm{s}}} = 1.93 \pm 0.28$ and $\frac{N_{x,\mathrm{L}}^{\mathrm{s}}}{N_{x,\sigma^-}^{\mathrm{s}}} = 1.88 \pm 0.26$. The saturation behaviour reflects the quenching of the excitonic oscillator strength at high excitonic densities. Under circularly-polarized excitation, both the saturation current and the characteristic saturation power decrease by nearly a factor of two compared to linear excitation, providing a quantitative fingerprint of the valley-selective control over many-body interactions.

**Spectral lineshape modulation at high exciton densities**

At elevated exciton densities, many-body interactions reshape the A-exciton resonance, producing characteristic fingerprints such as reduced overall intensity, blueshift of the peak energy, and linewidth broadening. We capture these effects by analysing the evolution of photocurrent spectra with exciton density for one- and two-valley excitation. Figures 2a, b show the photocurrent spectra at different exciton densities for linear (black) and $\sigma^+$ (red) excitation. For both polarizations, the excitonic resonance undergoes pronounced modifications. We first examine the device responsivity decrease around the A-exciton resonance. Figure 2c shows the photocurrent as a function of exciton density on a log–log scale. The solid lines correspond to fits of the experimental data using $I_{pc} \propto N_X^{\alpha}$. At low densities ($N_X < 10^{12}$ cm$^{-2}$), the photocurrent increases linearly with exciton density ($\alpha_{\sigma^+} = 0.97$ and $\alpha_L = 0.96$), but becomes sublinear ($\alpha_L = 0.77$ and $\alpha_{\sigma^+} = 0.7$), at higher densities ($N_X \approx 10^{12} - 10^{13}$ cm$^{-2}$), typical from exciton-exciton annihilation. To quantify the transition between these two regimes we obtain the intersection between the two fittings obtaining critical exciton densities $N_L = 1.09 \times 10^{12}$ cm$^{-2}$ and $N_{\sigma^+} = 0.51 \times 10^{12}$ cm$^{-2}$, for $L$ and $\sigma^+$ polarizations. The ratio of these crossover densities is roughly 2, indicating that single-valley excitation reaches the strongly interacting regime at roughly half of the density for linear excitation.

Next, we consider the exciton density-induced blueshift of the A-exciton. Figure 2d compares the blueshift of the peak position under $L$ and $\sigma^{+/-}$ excitation obtained from asymmetric Lorentzian fits (section S5), which accounts for the asymmetric peak broadening induced by exciton-exciton interactions[29]. At exciton densities of $1 \times 10^{13}$ cm$^{-2}$ we observe a blueshift of 28 meV for $\sigma^+$ excitation, whereas linear excitation yields only 8 meV, reflecting much stronger many-body interactions under single-valley pumping. Blueshifts of 10–20 meV reported in transient absorption experiments are typically attributed to phase-space filling and Hartree–Fock mean-field effects, both scaling with exciton density. Under linearly polarized excitation, however, the blueshift reaches only ~28% of the shift observed under circular excitation. A reduction of over a factor of 3 – rather than the factor of 2 expected from valley population alone – in the blueshift suggests a complex interplay of many-body exciton interactions. Because linear excitation populates both valleys, Coulomb repulsion and mean-field Hartree–Fock effects[32] are reduced relative to the single-valley population induced by circular excitation. Furthermore, the possible formation of bound biexcitons may further modify the spectral lineshape, reducing the blueshift[7]. This is consistent with coherent transient optical studies revealing biexcitonic fine structure in 1L-TMDs[33]. In an additional device we observe a secondary peak that can be attributed to biexciton formation (see section S6).

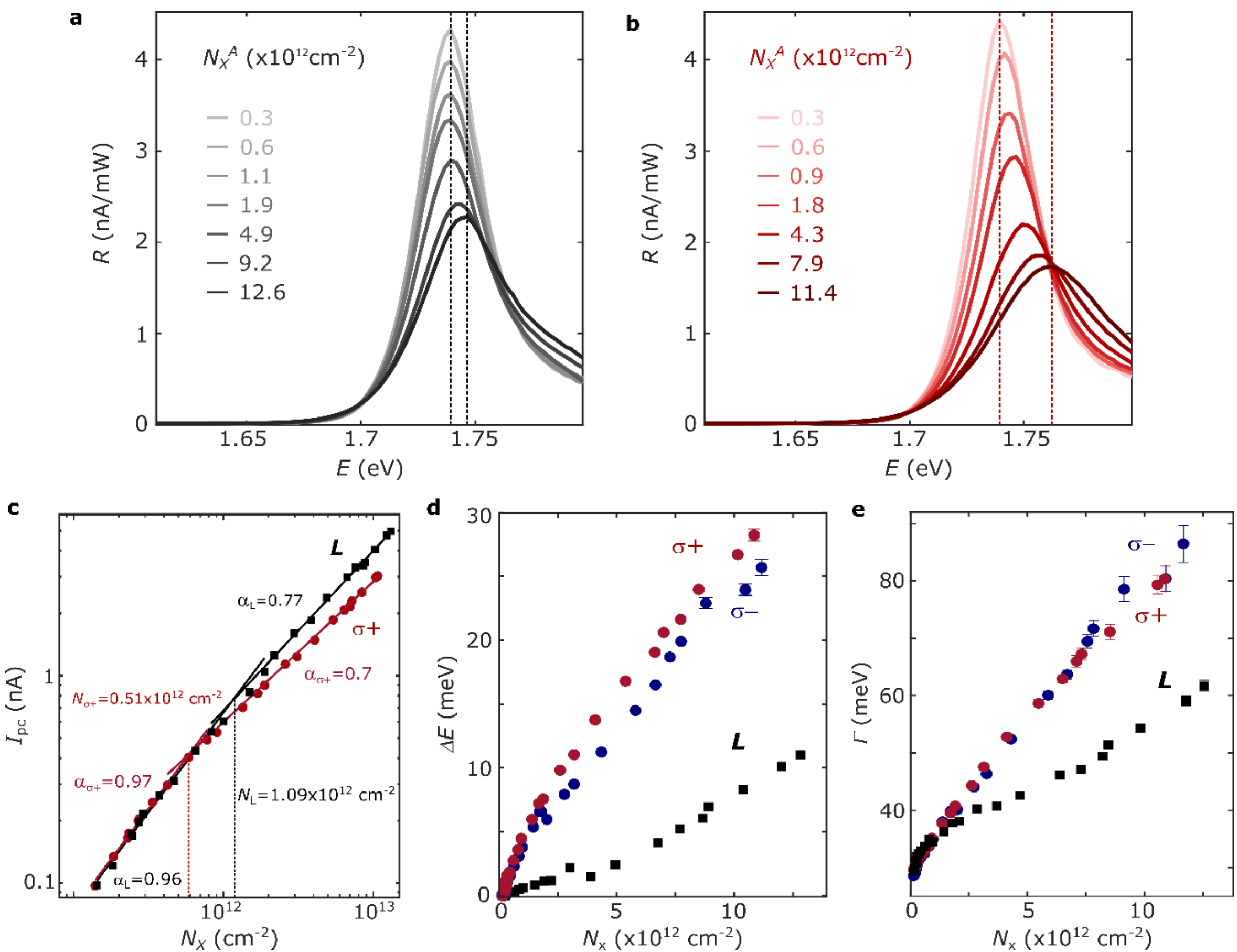

**Figure 2.** Exciton density dependence of A-exciton lineshape. (a, b) Responsivity spectra as a function of the exciton density for linear and $\sigma^+$ excitation. (c) Log–log plot of the photocurrent as a function of the exciton density for $L$ and $\sigma^+$ polarizations. The solid lines represent the fittings to $I_{pc} \propto N_X^{\alpha}$. The values of $\alpha$ and of the crossover densities are depicted in the graph. (d) Blueshift and (e) linewidth broadening of the A-exciton as a function of the exciton density for L and $\sigma^{+/-}$ excitation. The error bars correspond with the uncertainty in the fittings.

Linewidth broadening provides an additional fingerprint of many-body interactions (Fig. 2e). The A-exciton linewidth increases monotonically with exciton density. Under $\sigma^{+/-}$ excitation, the broadening exceeds 80 meV at the highest densities, compared with ~60 meV for linear excitation. This broadening reflects a reduced exciton lifetime arising from exciton-exciton annihilation, and excitation-induced dephasing due to exciton-exciton scattering[7,34,35]. The monotonic increase of the linewidth broadening points to many-body interactions as the dominant mechanism over inhomogeneous broadening[35]. The stronger broadening under circular excitation is consistent with the enhanced interaction strength when carriers are confined to a single valley. This contrast suggests that while the blueshift mainly encodes density-dependent renormalization of the exciton resonance, the linewidth provides a complementary measure of the lifetime reduction and exciton–exciton interactions.

**Temperature dependence and valley-controlled exciton-exciton annihilation.**

As mentioned before, several interaction channels can shape the nonlinear excitonic response in 1L-TMDs. As the exciton gas approaches the Mott transition, many-body effects become increasingly complex. Several mechanisms, such as phase-space filling, exciton-exciton scattering, Hartree-Fock exchange interactions or dynamical screening effects can also contribute to the nonlinearity of the photoresponse. Nonetheless, time-resolved photoluminescence[36], transient absorption[6] and photocurrent spectroscopy[37] studies have established that in the intermediate density regime relevant here ($10^{11} < N_x < 10^{13}$ cm$^{-2}$), exciton-exciton annihilation (EEA, or Auger-like exciton recombination) is the dominant decay channel for excitons in TMD monolayers, governing the dynamics in the equilibrium state which is the relevant framework for our measurements[38].

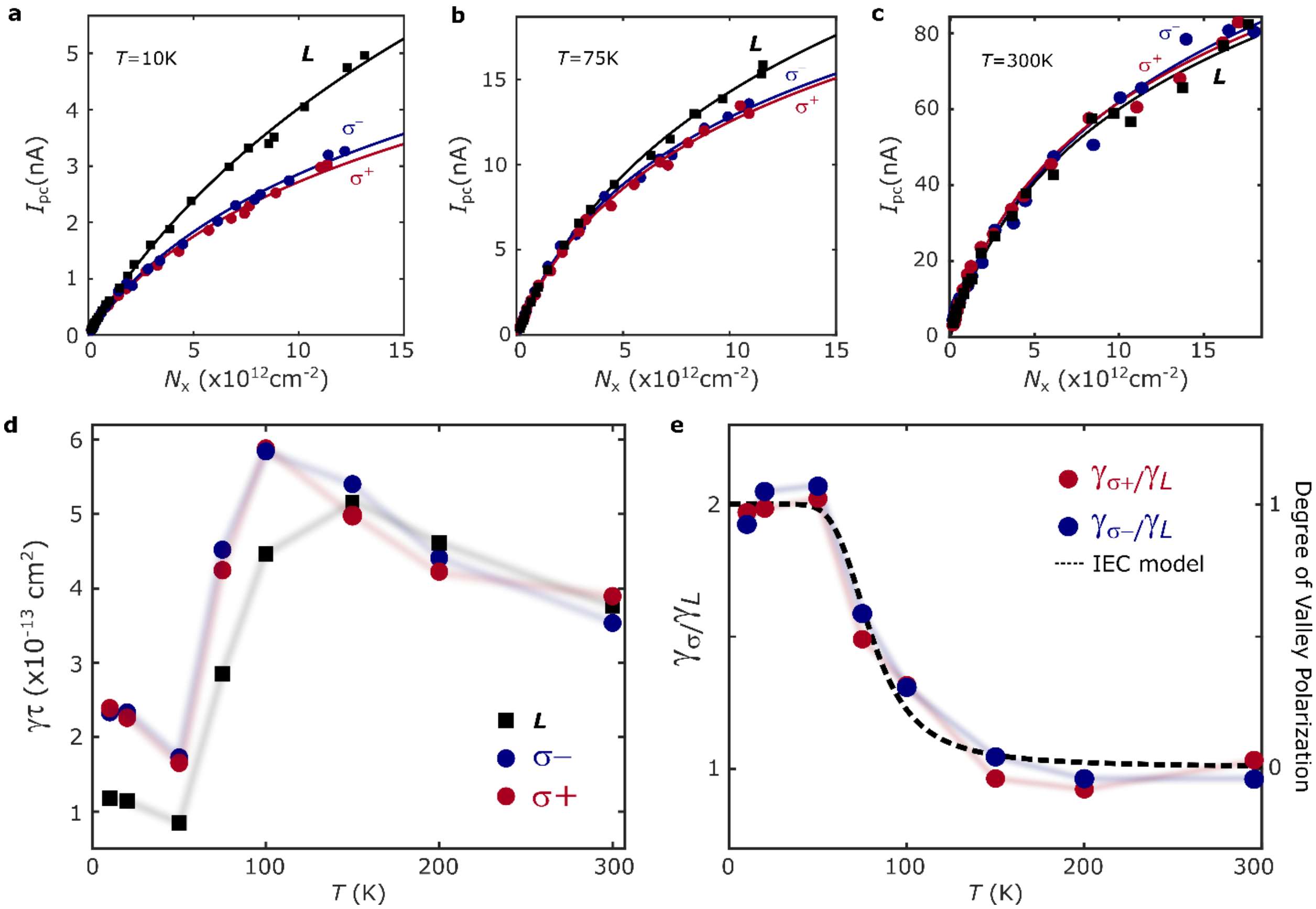

**Figure 3.** Temperature dependence of the valley-controlled exciton–exciton annihilation process. Photocurrent generated for linear and circular polarization as a function of the exciton density at temperatures of **(a)** 10 K, **(b)** 75 K, and **(c)** 300 K. The solid lines correspond with the fits to the exciton-exciton annihilation model $I_{pc} \propto \frac{1}{\gamma} ln(1 + \gamma\tau N_x)$. **d,** Temperature dependence of $\gamma\tau$ extracted from the fittings as a function of the temperature. **e,** Temperature dependence of the ratio between circular ($\sigma^+$ and $\sigma^-$) and linear (L) exciton-exciton annihilation coefficients (left-axis, dots) and of the degree of circular polarization calculated via the analytical model for intervalley exchange coupling (IEC) in 1L-$WSe_2$ (right-axis, dashed line).

To capture the influence of the Auger recombination, we use an equilibrium recombination model (see section S7 for more details) adding a quadratic loss term, $\gamma N_x^2$, in the exciton rate equation[37], where γ is the annihilation rate and *N* the exciton density. This model naturally yields a sublinear photocurrent response, which corresponds to:

$$I_{\mathrm{pc}} \propto \frac{1}{\gamma}\ln(1+\gamma\tau N_{\mathrm{x}}), \quad (2)$$

where $\tau$ is the photoresponse time of the device. To confirm the dominant role of exciton-exciton annihilation we perform nonlinear photocurrent measurements of the A-exciton under linear and circular excitation from cryogenic to room temperatures. In Figures 3a–c, we present the nonlinear photocurrent curves at 10 K, 75 K, and 300 K fitted using Equation (2). At 10 K, the contrast between linear and circular excitation is substantial, consistent with robust valley polarization. At 75 K, the difference is significantly reduced, and at room temperature, we observe a nearly identical behaviour for $L$ and $\sigma^{+/-}$ excitations, indicating a strongly reduced valley polarization and a reduction of valley-dependent many-body interactions. To account the spectral shift of the A-exciton peak as a function of the temperature we measure the complete photocurrent spectra at different temperatures (section S8). To get a better understanding of the photoresponse, we plot the extracted product of the exciton-exciton annihilation and the photoresponse time, $\gamma\tau$, of the device as a function of temperature (Fig 3d) for circularly- and linearly-polarized light excitation. It is important to note that for this analysis $\tau$ has been taken to be independent of the temperature, as supported by previous works[39]. Therefore, the temperature dependence is dominated by changes in $\gamma$. At temperatures below 50 K, the sublinearity of the photocurrent is less pronounced, resulting in $\gamma\tau$ of roughly $1 \times 10^{-13}$ $cm^2$. Above 50 K, $\gamma\tau$ dramatically rises by up to a factor of 5. This intriguing temperature dependence of the Auger recombination rate in 1L-$WSe_2$ has been demonstrated in a seminal study[27] combining theoretical microscopic calculations and photoluminescence measurements. It originates from the characteristic band structure of the 1L-$WSe_2$, in which the lowest-energy excitonic states are momentum-dark intervalley excitons composed of electrons located at the K' and Λ valleys and holes at K point of the valence band[40]. The rapid relaxation of bright excitons into energetically lower-lying dark states enables the equilibrium exciton population to be estimated by a Boltzmann distribution which upon integration over momentum gives the total exciton density. Under this description, the population is almost entirely governed by the intervalley KK' and KΛ dark excitons, which constitute the dominant excitonic species at thermal equilibrium (see section S9). Due to the energy splitting between these levels of around 10 meV[41], the Auger recombination rate is strongly dependent on the equilibrium thermal distribution of the excitonic states. Below 50 K, the equilibrium exciton density is predominantly composed of dark KK′ excitons, which exhibit reduced Auger recombination rates. However, at higher temperatures the thermal broadening activates the population of KΛ excitons, which undergo a more efficient exciton-exciton annihilation process, leading to a pronounced enhancement of the Auger recombination rate. The qualitative agreement between the microscopic calculations[27] and the experimental values of $\gamma\tau$ provides direct evidence that exciton-exciton annihilation constitutes the primary many-body interaction responsible for both the quenching of oscillator strength and the nonlinear scaling of the photocurrent in resonance with the A-exciton.

Having addressed the temperature-driven evolution of $\gamma\tau$ through the thermal occupation of dark and bright exciton states, we now analyse how the polarization-selective excitation modifies this behaviour. By extracting the ratios of the Auger coefficients, $\frac{\gamma_\sigma}{\gamma_L}$, as a function of the temperature, we directly probe the role of valley-selective excitation in the efficiency of exciton-exciton annihilation. Figure 3e shows the evolution of the experimentally-obtained $\frac{\gamma_{\sigma+}}{\gamma_L}$ and $\frac{\gamma_{\sigma-}}{\gamma_L}$ ratios as a function of temperature, obtaining similar results for $\sigma^+$ and $\sigma^-$ excitation. In the low-temperature limit, we obtain ratio of ~2 for both cases, as circular excitation injects twice as many excitons into a single valley compared with linear excitation. Building on our microscopic theory[27], the EEA coefficient is proportional to the square of the exciton population excited in a single valley, $\gamma_{\sigma+/\sigma-} \propto N^2_{\mathrm{KK/K'K'}}$. Hence, if half of the charge carriers are excited in each valley and taking the dominant bright exciton-exciton recombination channels into account we obtain $\gamma_L \propto \left(\frac{N_{\mathrm{KK}}}{2}\right)^2 + \left(\frac{N_{\mathrm{K'K'}}}{2}\right)^2 = \frac{1}{4}(\gamma_{\sigma+} + \gamma_{\sigma-}) = \frac{\gamma_\sigma}{2}$. This simple counting argument explains the factor-of-two difference observed experimentally below 50 K.

At elevated temperatures, the ratio of Auger coefficients decreases towards unity, reflecting a progressive valley depolarization of the exciton population. This trend arises from the activation of intervalley exchange coupling (IEC)[42] , which mediates spin and valley transfer between K and K' excitons as evidenced by microscopic theory and time-resolved ARPES studies[43] and strain dependent investigations[18]. The IEC is suppressed for momentum-forbidden dark excitons that dominate at low temperatures, but thermal occupation of bright states above ~50 K enables efficient intervalley exchange scattering. Beyond this threshold, valley polarization is only partially retained, and the contrast between circular and linear excitation diminishes. At higher temperatures, intervalley exchange fully redistributes the carriers between valleys, erasing the initial spin–valley polarization. To model this behaviour we start from the recent microscopic description of the degree of valley polarization (DOP) after continuous-wave excitation of strained monolayers[18] to describe our experiments with pulsed excitations. In particular, we develop an analytical model to calculate the DOP as a function of temperature (see section S10). In this model, the DOP is given by $DOP = \left(1 + \frac{\tau_{\mathrm{dec}}}{\tau_{\mathrm{spin}}}\right)^{-1}$, where $\tau_{\mathrm{spin}}$ is the spin relaxation time, , which we derive in a full microscopic way as $\tau_{\mathrm{spin}} = \frac{\hbar\Gamma_{\mathrm{x-ph}}}{4\langle|\mathrm{J}|^2\rangle}$, where $\Gamma_{\mathrm{x-ph}}$ is the thermally induced broadening and $|J|$ is the intervalley exchange interaction. The spin relaxation time is driven particularly by the average of the exchange interaction weighted by the relative occupation of KK states, with the momentum-dark excitons acting as a spin-preserving reservoir, and $\tau_{\mathrm{dec}}$ is a phenomenologically introduced recombination rate accounting for both radiative and non-radiative decay processes. As stated before, in 1L-TMDs spin and valley are intrinsically coupled, therefore $\tau_{\mathrm{spin}} = \tau_{\mathrm{valley}}$ [18]. On the right-axis of Figure 3e, we plot the DOP as a function of the temperature for $\tau_{dec} = 200$ ps. The qualitative agreement on the temperature dependence between the microscopical model and experiment confirms that intervalley exchange coupling of K-valley–polarized excitons and their scattering into the K'-valley (and vice-versa) are strongly suppressed at low temperatures, where excitons predominantly occupy momentum-dark states. Nevertheless, at temperatures above 50 K, the degree of polarization rapidly vanishes as thermal activation of bright states enhances exchange-induced scattering. This directly supports the assumption that the suppression of intervalley exchange scattering is the key mechanism sustaining long-lived valley polarization.

**Discussion**

In this work, we demonstrate the optical control of excitonic many-body interactions by exploiting the valley degree-of-freedom in 1L-$WSe_2$ devices through helicity-resolved photocurrent spectroscopy. We show that the main mechanism of exciton sublinearity, namely exciton-exciton annihilation can be actively controlled by valley selective excitation. This establishes a direct link between valleytronic control and excitonic many-body physics. Looking ahead, the ability to optically control many-body exciton interactions through the valley degree-of-freedom offers exciting opportunities for exploring and controlling correlated phenomena in engineered two-dimensional systems[21,44]. Furthermore, combining the valley-dependent measurements with electrostatic gating, promotes the formation of charged excitons and carrier-induced screening modifying the valley polarization [18,45], and offering new possibilities to control many-body exciton interactions. Extending this approach to moiré superlattices or hybrid heterostructures could provide access to exotic exciton phases such as exciton condensates at more accessible fluences and temperatures[46] while also offering an additional tuning knob on their interaction. This new route opens opportunities for developing excitonic, spintronic and valleytronic devices[16,47,48] that exploit optically-controlled nonlinearities in two-dimensional semiconductors, advancing the integration of many-body quantum phenomena into optoelectronic applications.

ASSOCIATED CONTENT

**Data availability statement**

The data that support this study are available at https://zenodo.org/records/19887546.

**Supporting Information**.

The Supporting Information is available free of charge at https://pubs.acs.org/doi/10.1021/acs.nanolett.6c01091.

(Section S1) Methods; (Section S2) Fabrication of the 1L-$WSe_2$/h-BN/graphite phototransistors; (Section S3) Low-temperature photocurrent spectroscopy setup; (Section S4) Transfer curves of the $WSe_2$ phototransistors; (Section S5) Photocurrent spectra fitting procedure; (Section S6) Additional device measurements; (Section S7) Exciton-exciton annihilation mediated photocurrent model; (Section S8) Temperature dependence of the A-exciton resonance; (Section S9) Occupation level of bright excitons KK and dark excitons KK' and KΛ; (Section S10) Microscopic model for temperature dependence of valley depolarization; (Section S11) Spatially-resolved photocurrent mappings.

AUTHOR INFORMATION

**Corresponding Authors**

Daniel Vaquero. Zernike Institute for Advanced Materials, University of Groningen, 9747 AG Groningen, The Netherlands. Email: d.vaquero.monte@rug.nl

Marcos H.D. Guimarães. Zernike Institute for Advanced Materials, University of Groningen, 9747 AG Groningen, The Netherlands. Email: m.h.guimaraes@rug.nl

**Author Contributions**

D.V. and M.H.D.G. conceived the idea for the experiment. D.V. performed the fabrication of monolayer $WSe_2$ phototransistors with the support of C.A.C.S. and M.H.T. The hBN crystals were provided by K.W. and T.T. D.V. and C.A.C.S. assembled and tested the optoelectronic setup. D.V., C.A.C.S., and M.H.T. carried out the photocurrent measurements. D.V. performed the data analysis. D.E., R.R. and E.M. developed the theoretical model for the temperature-dependent valley depolarization and provided the theoretical interpretation of the results. D.V. wrote the manuscript under the supervision of M.H.D.G. with the input from all the authors.

ACKNOWLEDGMENTS

The authors acknowledge J. G. Holstein, H. Adema, H. de Vries, A. Joshua, and F. H. van der Velde for their technical support. Sample fabrication and nanocharacterization were performed using the Zernike Nanolab facilities, which are part of the NanolabNL network.

This work was supported by the Zernike Institute for Advanced Materials, the research program "Materials for the Quantum Age" (QuMat, registration number 024.005.006), which is part of the Gravitation program financed by the Dutch Ministry of Education, Culture and Science (OCW), and the European Union (ERC, 2D-OPTOSPIN, 101076932). Views and opinions expressed are however those of the authors only and do not necessarily reflect those of the European Union or the European Research Council. Neither the European Union nor the granting authority can be held responsible for them. D.V. acknowledge funding from the project NWO XS "A New Route to Energy-Efficient Data Transfer: Spin–Light Interconversion in 2D Materials" (OCENW.XS25.2.170). D.E., R.R. and E.M. acknowledge funding from the Deutsche Forschungsgemeinschaft via the project 542873285 and from the LOEWE Exploration program of the State of Hesse, Germany. K.W. and T.T. acknowledge support from the JSPS KAKENHI (Grant Numbers 21H05233 and 23H02052), the CREST (JPMJCR24A5), JST and World Premier International Research Center Initiative (WPI), MEXT, Japan.

# Supporting Information

# Valley-Controlled many-body exciton interactions in monolayer $WSe_2$ phototransistors

Daniel Vaquero[1*], Cédric A. Cordero-Silis[1], Daniel Erkensten[2], Roberto Rosati[2], Martijn H. Takens[1], Kenji Watanabe[2], Takashi Taniguchi[3], Ermin Malic[2], Marcos H. D. Guimarães[1*]

*[1]Zernike Institute for Advanced Materials, University of Groningen, 9747 AG Groningen, The Netherlands*

*[2]Department of Physics, Philipps-Universität Marburg, 35037 Marburg, Germany.*

*[3]Research Center for Electronic and Optical Materials, National Institute for Materials Science, 1-1 Namiki, Tsukuba 305-0044, Japan*

*[4]Research Center for Materials Nanoarchitectonics, National Institute for Materials Science, 1-1 Namiki, Tsukuba 305-0044, Japan*

**email: d.vaquero.monte@rug.nl, m.h.guimaraes@rug.nl*

## S1: Methods

### Van der Waals heterostructure assembly and device fabrication

Thin flakes of graphite and hexagonal boron nitride (hBN) are mechanically exfoliated on oxidized silicon substrates (285 nm $SiO_2$/Si). The homogeneity and thickness of the flakes are identified by optical contrast. Graphite/hBN back gates are assembled by using a standard van der Waals dry pick-up[1]. The hBN and graphite flakes are subsequently picked up using a polycarbonate (PC)/polydimethylsiloxane (PDMS) stamp and placed on a clean substrate. Polymer residues remaining after the transfer process are removed using chloroform. Monolayer $WSe_2$ flakes were obtained by mechanical exfoliation from bulk crystals (HQ graphene) onto a PDMS substrate (GelFilm WFx4 × 6.0 mil by Gel-Pack). After optical identification, the monolayer $WSe_2$ is directly transferred onto the graphite/hBN stack[2]. The electrodes are fabricated using standard electron-beam lithography, reactive ion etching and electron-beam thin-film deposition of Pt (5 nm)/Au (50 nm).

### Photocurrent measurements

Photocurrent spectroscopy measurements were performed with the device mounted inside a Janis ST-500 continuous-flow cryostat under high vacuum (~5 x $10^{-7}$ mbar), with temperature control between 5 K and 300 K. The light source is a wavelength-tunable (690-1040 nm) 100 fs pulsed-laser based in a mode-locked Ti: Sapphire oscillator (Spectra-Physics MaiTai HP), with a repetition rate of 80 MHz. The laser beam was focused on the device using a 10x long-working-distance microscope objective (Mitutoyo), mounted on a XYZ positioning stage. To improve the stability of the measurements, the spot is expanded to a beam radius of $\sim 3.4$ µm. An optical chopper modulates the excitation light, and the electrical response of the device is registered using a lock-in amplifier with the same modulation frequency (177 Hz). All the measurements were performed at a bias voltage $V_{\mathrm{b}} = 5\ V$ and a gate voltage of $V_{\mathrm{g}} = 0\ V$. Additional details of the optoelectronic setup are provided in Supplementary Note 2.

## Microscopic modelling of circular polarization

To obtain microscopic access to the degree of circular polarization, we make use of a many-body Hamiltonian including a kinetic part and the intervalley exchange coupling (details in Supplementary Note 8). We derive the corresponding Heisenberg equations of motion for the momentum-integrated exciton occupations with opposite spin within the relaxation-time approximation and consider both bright and momentum-dark exciton species. The exciton landscape is obtained by solving the Wannier equation with material-specific parameters including carrier masses and valley splitting obtained from ab-initio calculations[3]. The degree of circular polarization is obtained by integrating the exciton occupations over time and determined by the ratio between the difference in spin-up and spin-down exciton densities, and the total exciton density.

## S2: Fabrication of the 1L-$WSe_2$/h-BN/graphite phototransistors.

The fabrication of the 1L-$WSe_2$/h-BN/graphite heterostructure is performed by standard van der Waals heterostructure assembly methods[4]. Figure S1 summarizes the fabrication steps of the device. The few-layer h-BN (NIMS) and graphite (HQ graphene) flakes are mechanically exfoliated through micromechanical cleavage on a $SiO_2$ (285 nm)/Si wafer. The 1L-$WSe_2$ is exfoliated using polydimethylsiloxane (PDMS – Gel-Pak WF-X4) as a viscoelastic substrate. The flakes are identified through optical contrast and the 1L-$WSe_2$ is characterized by the use of the RGB channels[5] and with an accumulated photoluminescence (PL) microscope, see Fig S1b. After the exfoliation, the h-BN and graphite flakes are stacked together by using a transfer process using polycarbonate (PC)/PDMS stamps[1] (see Fig S1a). The h-BN and graphite are subsequently picked up using the PC/PDMS stamp at temperatures between $40^{o}$ C and $60^{o}$ C and placed on top of the $SiO_2$/Si substrate by melting the PC stamp at $200^{o}$ C. Finally, the heterostructure is cleaned in a chloroform bath and rinsed with acetone and isopropyl alcohol to remove the PC residues. Subsequently, the 1L-$WSe_2$ is transferred onto the heterostructure from the PDMS[2].

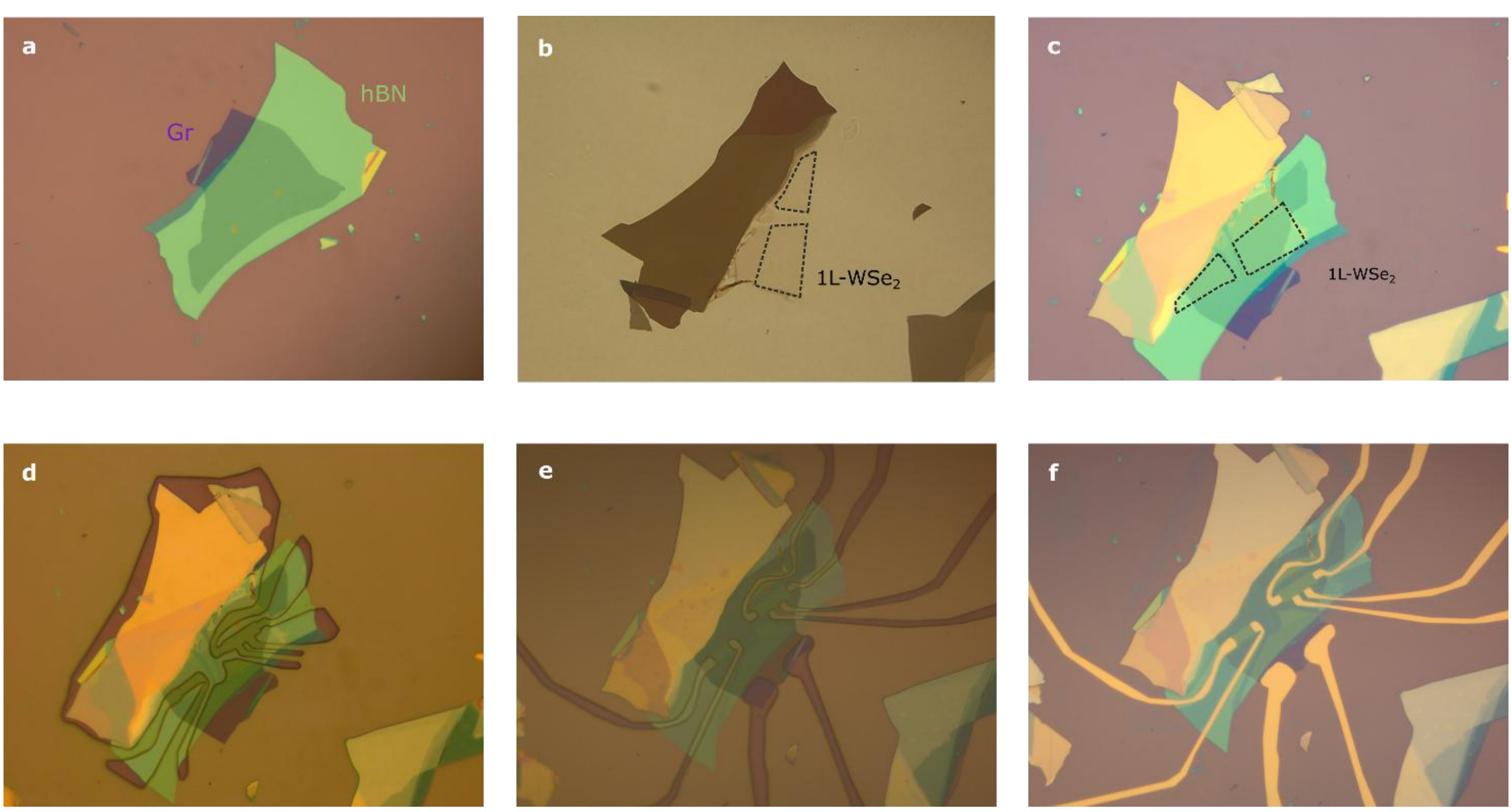

**Figure S1.** Fabrication process of the 1L-$WSe_2$/hBN/Gr device. (a) Assembled bottom gate of graphite/h-BN. (b) Exfoliated 1L-$WSe_2$ on PDMS (Gel-Pak WF-X4). (c) Final heterostructure after the dry transfer of 1L-$WSe_2$ on top of the graphite/h-BN. (d) PMMA mask for the definition of the device geometry. (e) Second lithography process for the definition of the electrical contacts. (f) Final device after the e-beam evaporation of the platinum/gold contacts and the lift off process.

After assembling the graphite/hBN/1L-$WSe_2$ heterostructure, the device geometry is defined using electron-beam lithography (EBL) on a Raith e-Line Plus system. A bilayer polymethylmethacrylate (PMMA) resist (50K and 950K, 9% wt.) is employed. Each PMMA layer is spin-coated at 4000 rpm for 1 minute, and baked at 180 °C for 40 s. The resist is then exposed with an electron dose of 170 $\mu C\ cm^{-2}$ at 10 kV and developed in a 1:3 MIBK:isopropanol solution. The resulting patterned regions are used to define the device geometry (see Fig. S1d).

Next, the exposed areas are etched by reactive ion etching (RIE) in an $CHF_4$ atmosphere using an ICP-RIE. After etching, the sample is thoroughly cleaned in acetone and isopropanol. Following geometry definition, a second EBL step is performed to pattern the metallic contacts (Fig. S1e). Special care is taken during electrode design to ensure clean alignment and to prevent overlap with multilayer $WSe_2$ regions.

Finally, platinum/gold (5/50 nm) contacts are deposited by electron-beam evaporation under high-vacuum conditions ($\sim 10^{-7}$ mbar). The fabrication concludes with a lift-off process in hot acetone, completing the final device shown in Fig. S1f.

## S3: Low-temperature photocurrent spectroscopy setup

The experimental setup for photocurrent spectroscopy is schematically depicted in Supplementary Figure S2. The sample is mounted inside a Janis ST-500 continuous flow cryostat under high vacuum (~$10^{-7}$ mbar), with temperature control ranging from 5 K to 300 K. The light source is a ~100-fs-long laser pulses generated by a mode-locked Ti:Sapphire oscillator (Spectra-Physics MaiTai), at a repetition rate of 80 MHz. The spectral range is 690-1040 nm allowing to scan the NIR spectral range. The scan of the wavelengths is performed in steps of 1-2 nm and the pulse width is between 5 and 10 nm. To monitor the illumination power throughout our measurements we use a silicon photodetector (Thorlabs PD36A2) with a response flattener filter (Thorlabs SRF11) which has been calibrated using an additional power meter.

The light polarization is controlled using a linear polarizer and an achromatic quarter-waveplate placed in a rotation mount. The laser beam was first collimated and then focused on the crystals using a 10x microscope objective placed on a XYZ positioning stage with micrometric resolution. The spot radius is adjusted to ~3.5 μm in order to avoid instabilities in the measurement. The setup also includes a halogen lamp and a CCD camera, aligned with the laser which allows for an easy sample alignment. The excitation light is modulated by an optical chopper and the electrical response of the device is registered using a lock-in amplifier with the same modulation frequency (177 Hz).

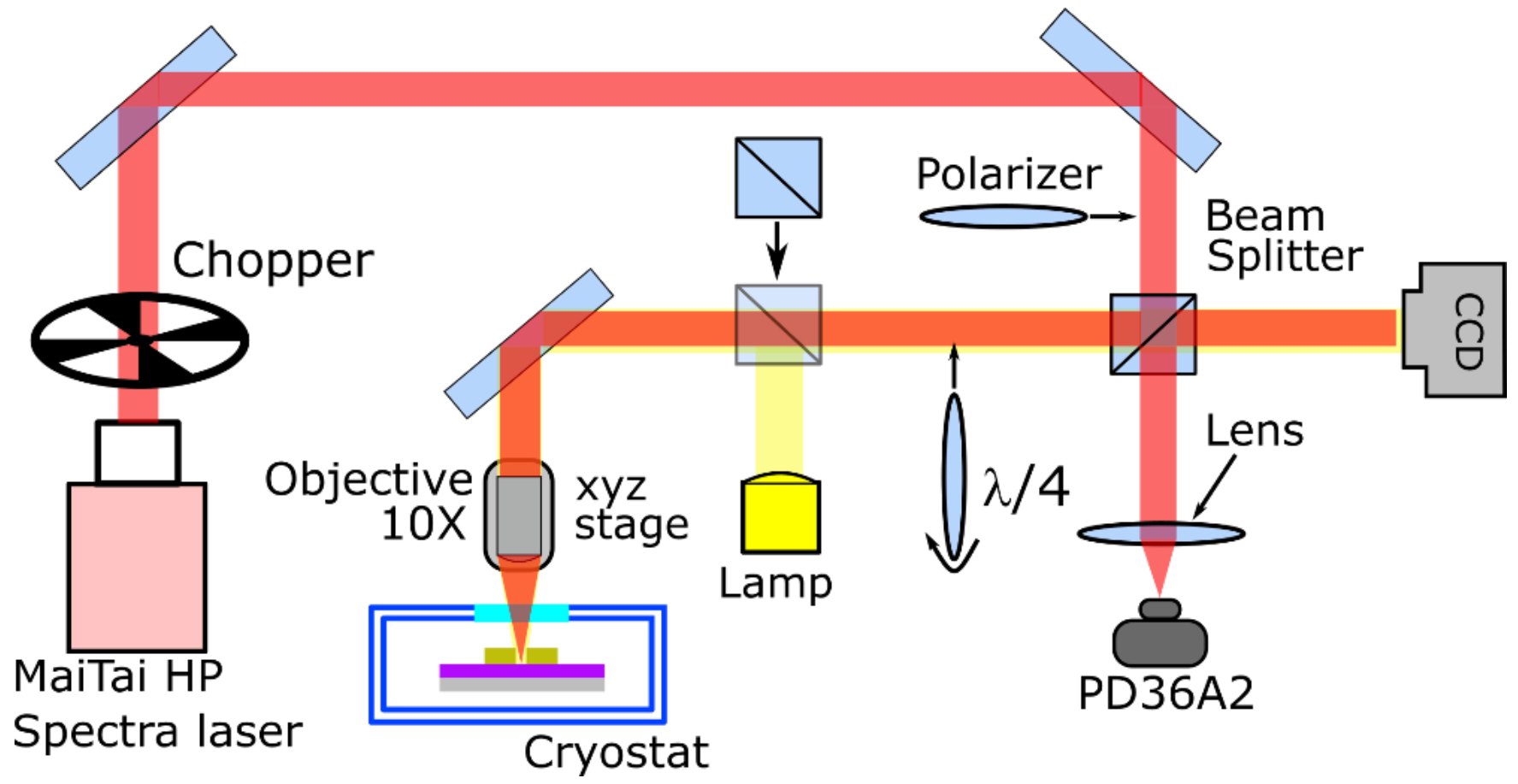

**Figure S2.** Schematic of the low temperature photocurrent

## S4: Gate transfer curves of the $WSe_2$ phototransistors.

Figure S3 shows the electrical transfer curves of our monolayer $WSe_2$ phototransistors. Both devices exhibit an *n*-type transistor behaviour. The electrical conductivity of the semiconductor channel increases with gate voltage as the Fermi level shifts into the conduction band, thus increasing the free-carrier density. The threshold voltage of the device is calculated by fitting the linear region of $I_{ds}$ as a function of $V_g$ (red dashed lines in Figure S3), obtaining a threshold voltage of $V_{th} = 5.3$ V and $V_{th} = 4.5$ V for devices D1 and D2, respectively.

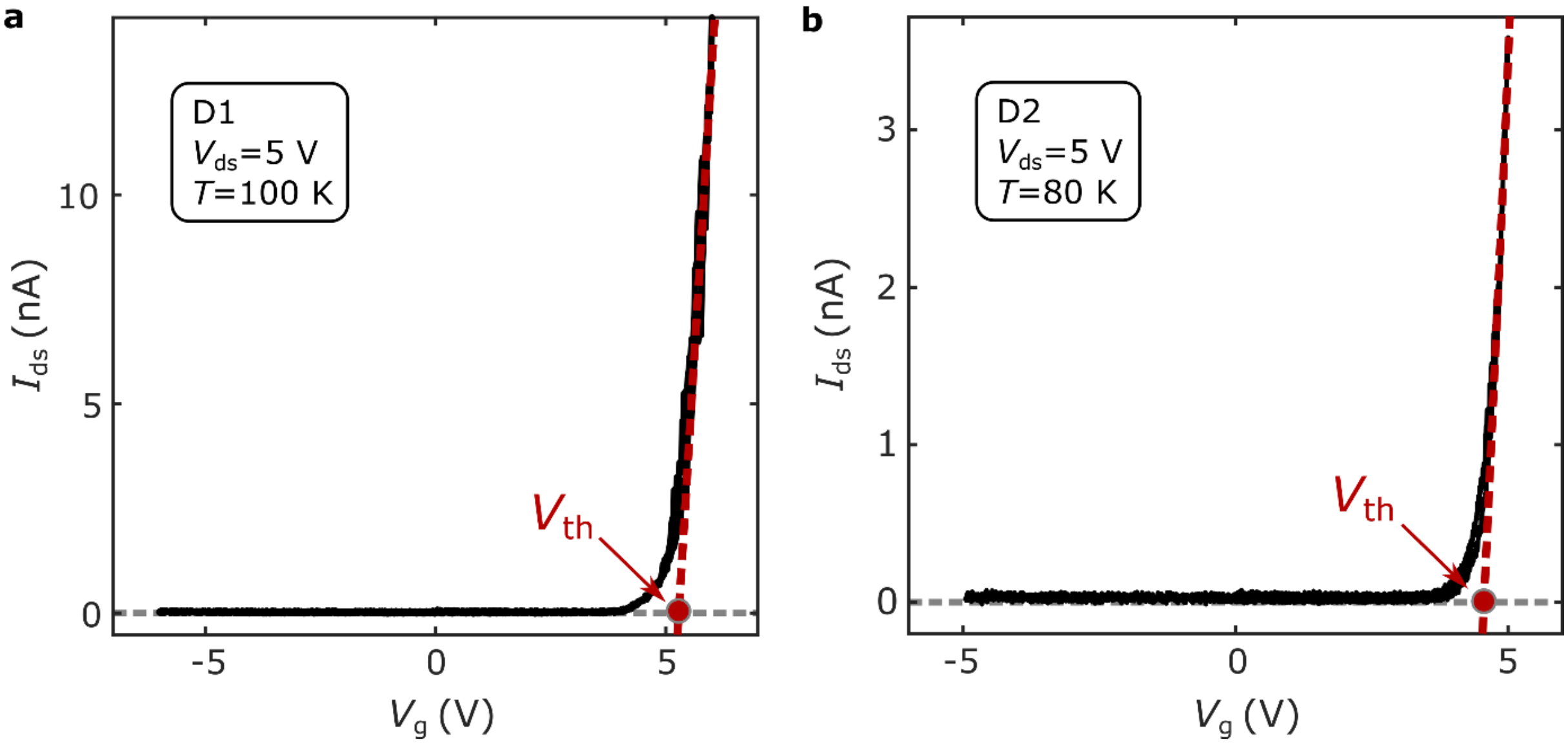

**Figure S3.** Gate transfer characteristics of the monolayer $WSe_2$ phototransistors. $I_{ds}$ as a function of $V_g$ measured at $V_{ds} = 5$ V for device D1 at $T = 100$ K (a) and device D2 at $T = 80$ K (b). The dashed red lines indicate linear fittings used to extract the threshold voltages, obtaining $V_{th} = 5.3$ V for device D1 and $V_{th} = 4.5$ V for device D2.

.

## S5: Photocurrent spectra fitting procedure

To fit the low-temperature photocurrent spectra of the 1L-$WSe_2$ device at different fluences, we use a modified Lorentzian function with an asymmetric profile. Under resonant excitation of the A-exciton, it is expected that the spectral lineshape of the A-exciton presents an asymmetric profile, which can be generated by scattering-induced dephasing phenomena[6] or dynamical screening effects[7]. This function consists in a regular Lorentzian function with an energy-dependent broadening[8] :

$$I(E) = \frac{I_0}{\gamma_0} \frac{1 + e^{a(E-E_0)}}{1 + \left( \frac{E - E_0}{\frac{\gamma_0}{1 + e^{a(E-E_0)}}} \right)^2}, \tag{S1}$$

where $I_0$ is the oscillator strength, $\gamma_0$ is the static broadening or full-width at half maximum (FWHM), $a$ is the asymmetry parameter and $E_0$ is the center of the lorentzian peak. Figures S3 shows the responsivity spectra measured under different illumination fluences. The low fluence (2 μJ/cm$^2$) spectrum is shown in Fig. S4a, while Fig. S4b and c show the spectra measured at higher fluences (75 μJ/cm$^2$) for linear and circular polarizations, respectively. All the spectra are fitted using Equation S1. The asymmetric Lorentzian function captures the spectral lineshape of the A-exciton peak with a good agreement between the fitting and the experimental data under different excitation conditions.

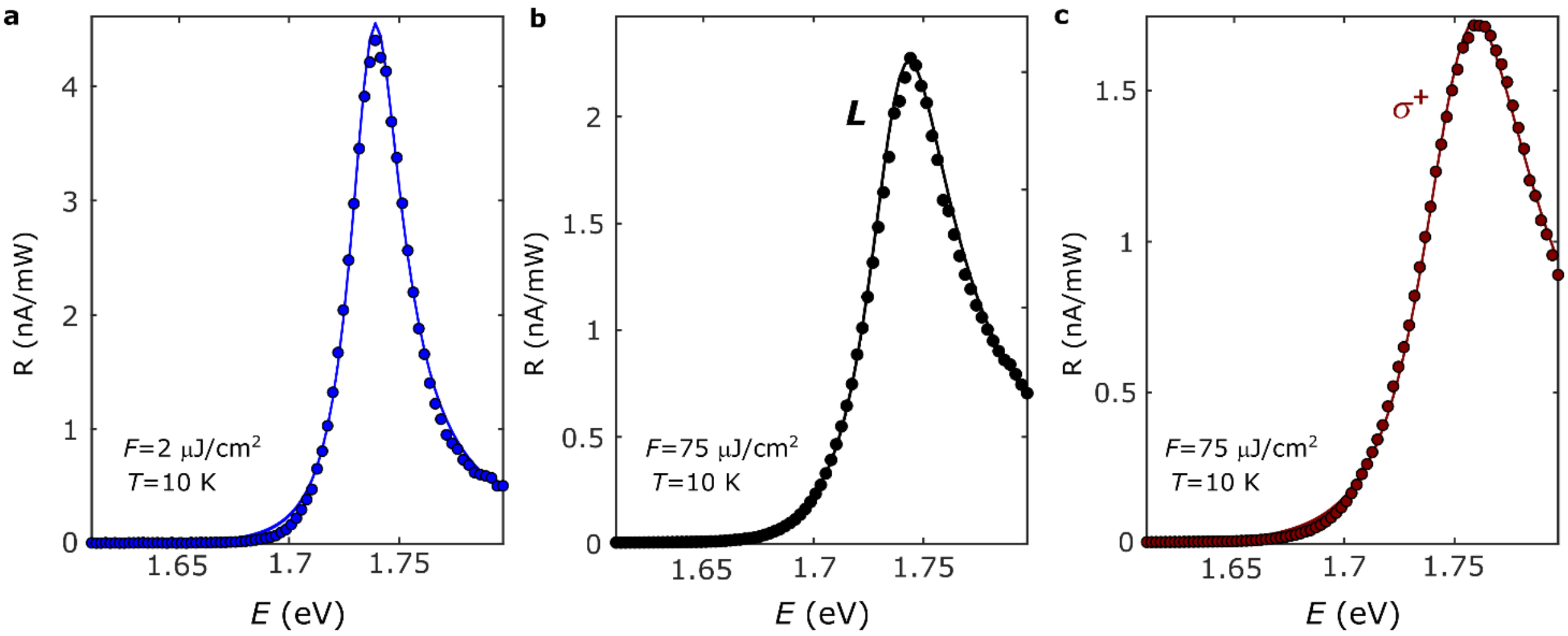

**Figure S4.** Asymmetric Lorentzian function fitting of the photocurrent spectra. Responsivity spectra measured under (a) low fluence (blue) and (b) high fluence linear polarization (black), and (c) high fluence circular polarization (red), respectively. The dots correspond to the experimental data and the solid lines to the fittings to Equation S1.

## S6: Additional device measurements.

To check for reproducibility of our results, we measured another device (D2) at liquid nitrogen temperatures, $T$ = 80 K. All the measurements presented here were performed at similar bias conditions than those shown in the main text for device D1, $V_{\mathrm{ds}} = 5\,V$ and $V_{\mathrm{g}} = 0\,V$. Figure S5a shows the optical image of the device D2. The architecture of the device follows a similar geometry to that of device D1 shown in the main text, with a graphite/hBN/1L-$WSe_2$ architecture and Pt/Au contacts. Figures S5b, c show the photocurrent spectra at different exciton densities for linear (black) and $\sigma^+$ (red) excitation. At low exciton densities, the photocurrent spectra are dominated by the strong

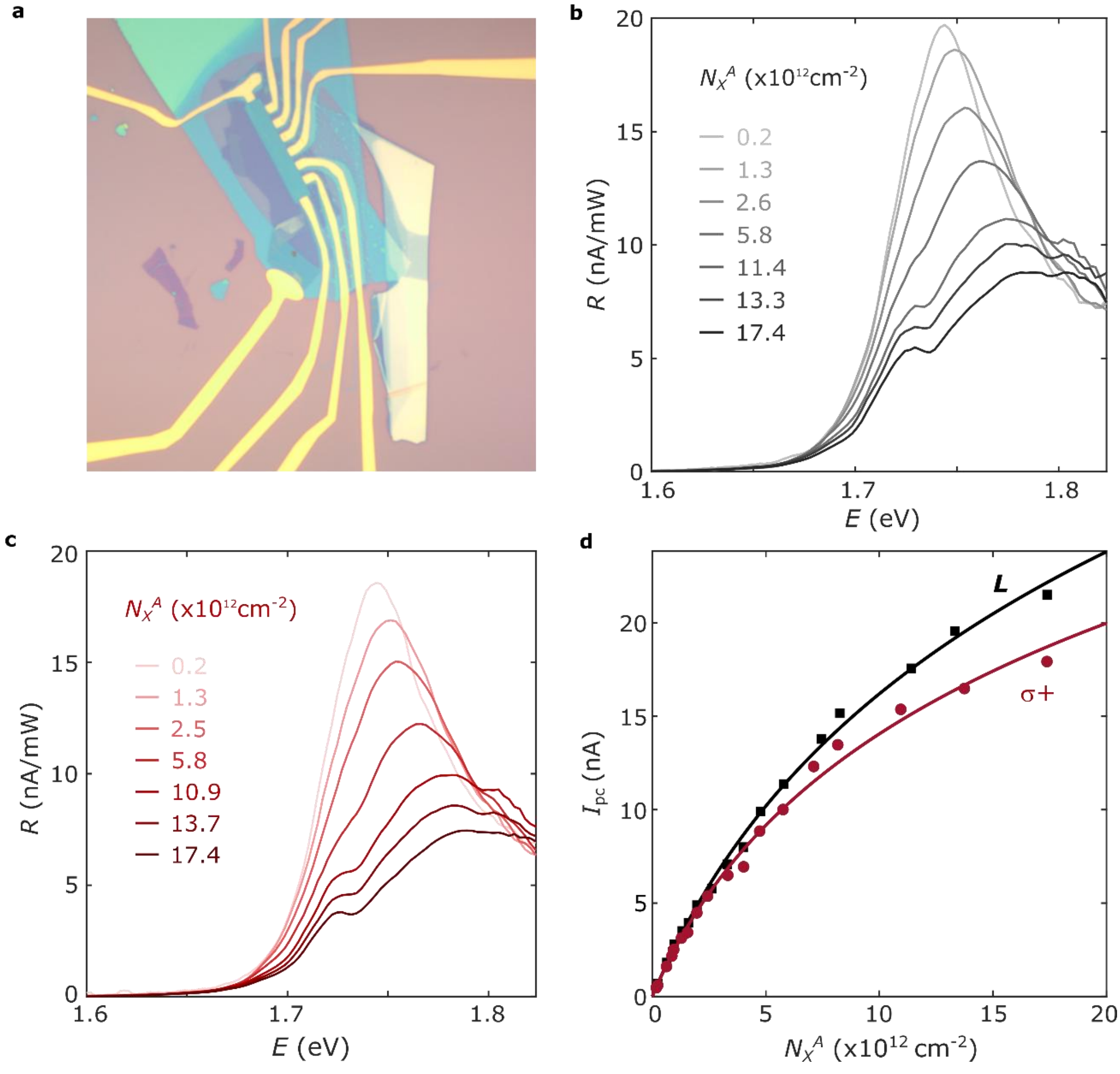

**Figure S5.** Device D2 measurements at $T$ = 80 K. (a) Optical image of the device D2. Responsivity spectra as a function of the exciton density for (b) linear and (c) $\sigma^+$ excitation. (d) Photocurrent generated for linear and circular polarization as a function of the exciton density. The solid lines correspond with the fits to the exciton-exciton annihilation model $I_{pc} \propto \frac{1}{\gamma} ln(1 + \gamma\tau N_x)$.

A-exciton resonance peak. Similarly to device D1, we observe a reduction of the device responsivity as the exciton density increases. Furthermore, at high exciton densities, $N_X^A > 10^{13}$ cm$^{-2}$, a secondary peak appears in the photocurrent spectra for linear and circular polarization. This peak could be related to different excitonic species, such as biexciton or trion formation. Following a similar procedure to that in the main text, we analyse the sublinearity of the photocurrent generated by the increase of the exciton density in resonance with the A-exciton. In Figure S5d, we present the sublinear photocurrent behaviour as a function of the exciton density at a laser energy resonant to the A-exciton. We observe a stronger sublinearity for circular excitation than for linear, confirming the experimental observations for device D1. The solid lines correspond with the fitting of the photocurrent dependence with the exciton density to the exciton–exciton annihilation sublinear model (Equation S4). From the fittings, we extract values of $\gamma_{\sigma^+}\tau = 2.66 \times 10^{-13}$ cm$^{-2}$ and $\gamma_L\tau = 1.95 \times 10^{-13}$ cm$^{-2}$, obtaining a ratio of $\frac{\gamma_{\sigma+}}{\gamma_L} = 1.37$, in line with those obtained for device D1.

## S7: Exciton-exciton annihilation mediated photocurrent model

The sublinear dependence of the photocurrent as a function of the power density can be modelled by an exciton generation rate equation including the exciton–exciton annihilation loss term that depends quadratically with the exciton density:

$$\frac{dN_x}{dt} = G(t) - \frac{N_x}{\tau} - \gamma N_x^2 \, . \tag{S2}$$

Here, $N_x$ is the exciton density, $G(t)$ is the time-dependent generation rate, $\tau$ represents the characteristic photoresponse time, and $\gamma$ is the exciton–exciton annihilation coefficient. For pulsed excitation, the generation rate can be written as $G(t) = N_0\delta(t)$ with $N_0$ the initialized exciton density. Under this optical excitation conditions Equation S2 is analytically solvable obtaining:

$$N_x(t, N_0) = \frac{N_0 \, e^{-t/\tau}}{1 + \gamma\tau N_0(1 - e^{-t/\tau})} \, . \tag{S3}$$

The measured photocurrent under steady state conditions is proportional to the time integrated exciton population:

$$I_{PC}(N_0) \propto \int_0^{\infty} N_x(t, N_0) \, \mathrm{d}t = \frac{1}{\gamma} \ln(1 + \gamma\tau N_0) \, . \tag{S4}$$

Equation S4 effectively captures the observed sublinear dependence of the photocurrent on excitation power (see Fig. 3 in the main text).

To connect the photocurrent with the optical excitation conditions, the initial exciton density can be expressed as:

$$N_0 = \frac{F\alpha}{E_{ph}} \, , \tag{S5}$$

where $F$ is the excitation fluence, $\alpha$ is the absorption coefficient that is estimated to be 5% for 1L-$WSe_2$ stacked on h-BN [9] and $E_{ph}$ is the excitation photon energy. By fitting Eq. S4 to the measured exciton-density dependence, we extract the product $\gamma\tau$ as a function of the temperature, as presented in Fig. 3d of the main text.

## S8: Temperature dependence of the A-exciton resonance

To complement the analysis of the evolution of the photocurrent spectra with temperature, we characterize the spectral position as a function of the A-exciton as a function of the temperature. From this evolution, we can extract the characteristic energy of the phonon that couples with the exciton. The evolution of the exciton energy can be tracked by the use of the equation [10,11]:

$$E_g(T) = E_0 - \frac{2S\langle\hbar\omega\rangle}{e^{\frac{\langle\hbar\omega\rangle}{k_B T}} - 1} , \qquad \text{(S6)}$$

where, $E_g$ $(T)$ is the value of the optical band gap of the 1L-$WSe_2$, $S$ corresponds to the electron-phonon coupling energy, and $\hbar\omega$ represents the characteristic phonon energy. Figure S6a shows the normalized responsivity spectra measured in the low-density regime, $N_X < 3 \times 10^{11}$ cm$^2$. Figure S6b shows the spectral position of the A-exciton as a function of temperature. From the fittings we extract the value a phonon characteristic energy of 16.6±0.8 meV in good agreement with previous literature [12,13].

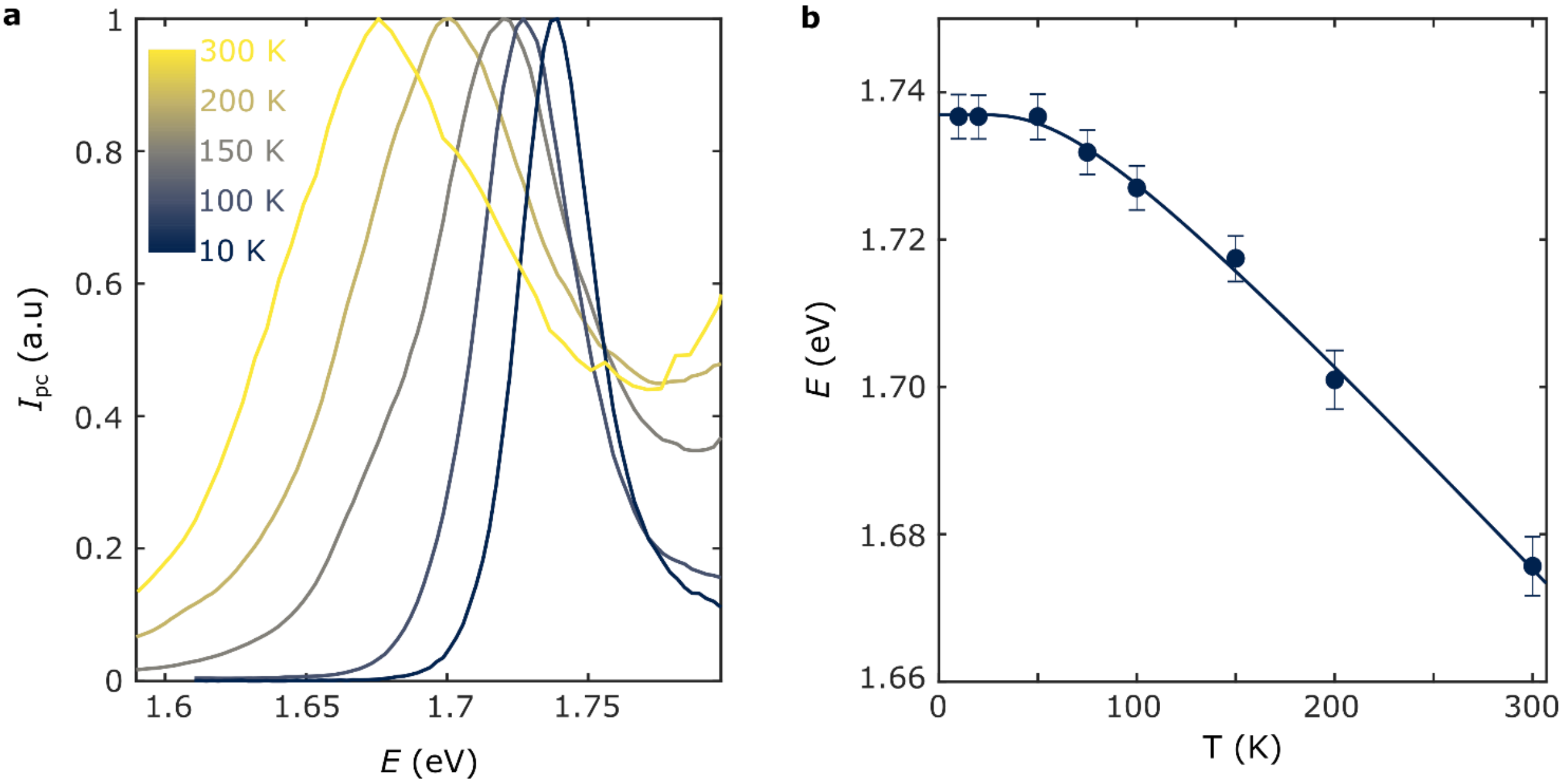

**Figure S6. Temperature evolution of the spectral position of the A-exciton peak.** (a) Normalized responsivity spectra measured at temperatures of 10 K, 100 K, 150 K, 200 K and 300 K. (b) Spectral position of the A-exciton peak. The error bars correspond to the wavelength step of 1-2 nm in the responsivity spectra. The solid line corresponds to the fit using Equation S6.

## S9: Occupation level of bright excitons KK and dark excitons KK' and KΛ

The band structure of the monolayer $WSe_2$ reflects the dark nature of the equilibrium excitons with the K' and the Λ point below the bright excitons[14]. Figure S7a sketches the band structure of a 1L-$WSe_2$ with the KK' and KΛ valleys, 46 and 34 meV below the bright exciton KK[15]. To capture the exciton occupation as function shows the exciton occupation of KK, KK' and KΛ levels as a function of the temperature calculated by the thermal Boltzmann distribution.

Figure S7b shows the occupation fraction of excitons as a function of the temperature estimated by using a thermalized Boltzmann distribution. The effective masses used for the calculation are obtained from Table VI in Ref. 16. The simulated occupation shows the dominance of dark KK' excitons at low temperatures. Over 50 K the KΛ excitons start to activate. Due to the three-fold degeneracy of the Λ-valley, the KΛ excitons shows a higher occupation number.

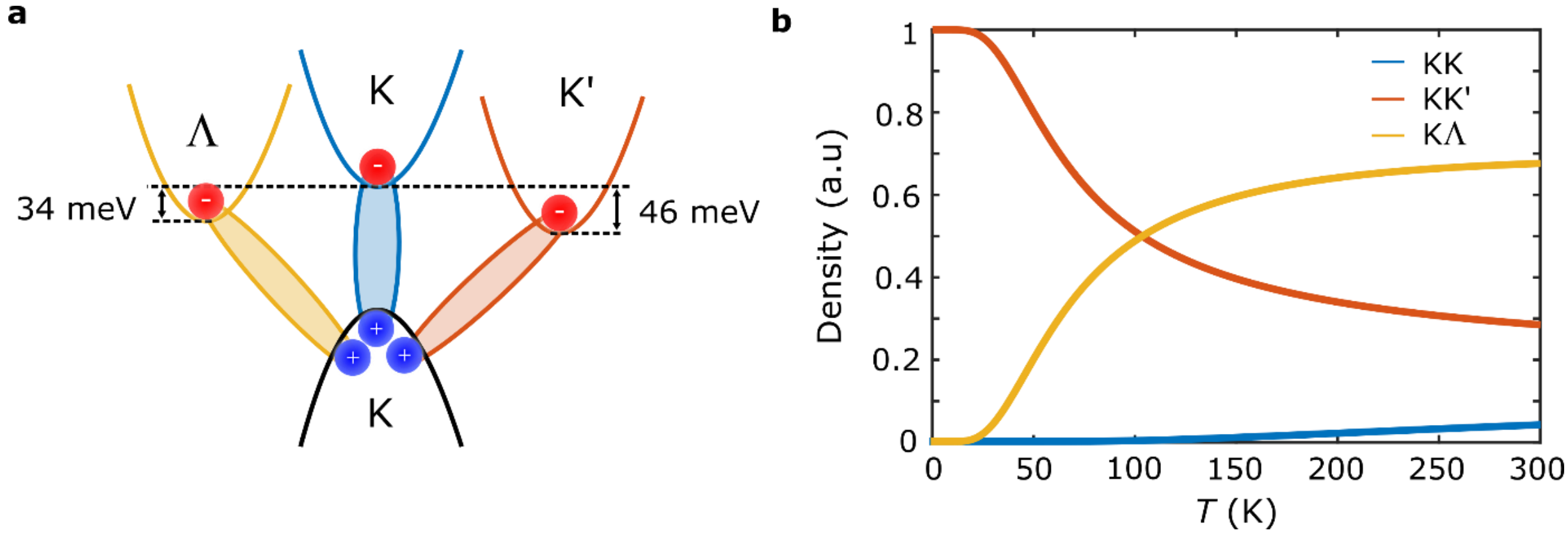

**Figure S7.** Occupation level of the bright and dark excitons in 1L-$WSe_2$. (a) Schematic of the three excitonic level system with bright exciton KK depicted in blue and the dark excitons, KK' and KΛ, depicted in orange and yellow respectively. (b) Valley occupation of the excitonic states KK (blue), KK' (orange) and KΛ as a function of the temperature.

## S10: Microscopic model for temperature dependence of valley depolarization

To model the valley depolarization, we start from the equations of motion for the (incoherent) exciton occupation $N^{s}_{\nu,\boldsymbol{Q}} = \langle X^{\dagger s}_{\nu,\boldsymbol{Q}} X^{s}_{\nu,\boldsymbol{Q}} \rangle$ as derived previously[16]:

$$\partial_t \, N^{s}_{\nu,\boldsymbol{Q}} = \partial_t N^{s}_{\nu,\boldsymbol{Q}}|_{\text{scat.}} + \frac{2}{\hbar}\mathrm{Im}\big(J^{s}_{\boldsymbol{Q}} C^{ss'}_{\nu_b,\boldsymbol{Q}}\big)\big(1-\delta_{s,s'}\big) - \frac{N^{s}_{\nu,\boldsymbol{Q}}}{\tau_{dec}}, \tag{S7}$$

where $\partial_t N^{s}_{\nu,\boldsymbol{Q}}|_{\text{scat.}}$ contains contributions due to exciton-phonon scattering, $J^{s}_{\boldsymbol{Q}}$ is the intervalley exchange interaction, and $\tau_{dec}$ is a phenomenologically introduced momentum-independent decay time taking into account radiative as well as non-radiative recombination due to e.g. defect-assisted scattering or spin-flip processes[16]. Furthermore, the equation depends on the exciton spin $s = \uparrow\uparrow, \downarrow\downarrow$ exciton valley $\nu = \mathrm{KK}, \mathrm{KK}', \mathrm{K\Lambda} \ldots$ (first letter denoting the valley of the hole and the second letter denoting the valley of the electron) and the center-of-mass momentum $\boldsymbol{Q}$. The equation of motion for the exciton occupation couples to the equation of motion for the transition correlation $C^{ss'}_{\nu_b,\boldsymbol{Q}} = \langle X^{\dagger s}_{\nu_b,\boldsymbol{Q}} X^{s'}_{\nu_b,\boldsymbol{Q}} \rangle$:

$$\partial_t \, C^{ss'}_{\nu_b,\boldsymbol{Q}} = \partial_t C^{ss'}_{\nu_b,\boldsymbol{Q}}|_{\text{scat.}} + \frac{1}{i\hbar}\big(J^{s}_{\boldsymbol{Q}}\big)^{*}\big(N^{s}_{\nu_b,\boldsymbol{Q}} - N^{s}_{\bar{\nu}_b,\boldsymbol{Q}}\big). \tag{S8}$$

Here, $\nu_b = \mathrm{KK}, \mathrm{K'K'}$, i.e. a bright valley and $\bar{\nu}_b$ denotes the opposite valley of $\nu_b$, i.e. if $\nu_b = \mathrm{KK}$, $\bar{\nu}_b = K'K'$ and vice versa. Next, analogously to the recent derivation for valley polarization after continous-wave excitation in strained monolayers[17] we write the transition correlation as a sum of the equilibrium (eq.) correlation and a deviation from equilibrium:

$$C^{ss'}_{\nu_b,\boldsymbol{Q}} = \delta C^{ss'}_{\nu_b,\boldsymbol{Q}} + \big(C^{ss'}_{\nu_b,\boldsymbol{Q}}\big)^{eq.} \,. \tag{S9}$$

Then, we solve the equation of motion for $C$ within the relaxation time approximation:

$$\delta \, C^{ss'}_{\boldsymbol{Q}} = \frac{\big(\mathrm{J}^{\mathrm{s}}_{\mathrm{Q}}\big)^{*}}{\mathrm{i}\Gamma_{\mathrm{x-ph}}}\big(\mathrm{N}^{\mathrm{s}}_{\mathrm{v_b},\boldsymbol{Q}} - \mathrm{N}^{\mathrm{s}'}_{\bar{\mathrm{v}}_{\mathrm{b}},\boldsymbol{Q}}\big)\,, \tag{S10}$$

where $\Gamma_{x-ph} = \sum_{\nu,\boldsymbol{Q}} \Gamma_{\mathrm{KK}\to\nu,\boldsymbol{Q}}$ is the phonon-induced linewidth. By plugging in the solution above into Eq. S7, we obtain

$$\partial_t N^s_{\nu,\boldsymbol{Q}} = \partial_t N^s_{\nu,\boldsymbol{Q}}|_{\text{scat.}} + \frac{2}{\hbar\Gamma_{\text{x-ph}}}\left|J^s_{\boldsymbol{Q}}\right|^2\left(\mathrm{N}^s_{\mathrm{v_b},\boldsymbol{Q}} - \mathrm{N}^{s'}_{\bar{\mathrm{v}}_\mathrm{b},\boldsymbol{Q}}\right)\left(1-\delta_{s,s'}\right) - \frac{N^s_{\nu,\boldsymbol{Q}}}{\tau_{dec}}. \tag{S11}$$

We sum over momentum $\boldsymbol{Q}$ and exciton states ν and fix, without loss of generality, the spin $s=\uparrow\uparrow$ resulting in

$$\partial_t N^{\uparrow\uparrow} = -\frac{2}{\hbar\Gamma_{\text{x-ph}}}\left(\langle|J^{\uparrow\uparrow}|^2\rangle_{\uparrow\uparrow}N^{\uparrow\uparrow} - \langle\left|J^{\uparrow\uparrow}\right|^2\rangle_{\downarrow\downarrow}N^{\downarrow\downarrow}\right) - \frac{N^{\uparrow\uparrow}}{\tau_{dec}}\ , \tag{S12}$$

with $N^s = \sum_{\nu,\boldsymbol{Q}} N^s_{\nu,\boldsymbol{Q}}$ and introducing the thermally averaged intervalley exchange interaction,

$$\langle|J^s|^2\rangle_{s'} = \frac{1}{N^{s'}}\sum_{\nu,\boldsymbol{Q}}\left|J^s_{\boldsymbol{Q}}\right|^2 N^{s'}_{\nu,\boldsymbol{Q}}. \tag{S13}$$

Next, we assume that $N^{\uparrow\uparrow}_{\nu,\boldsymbol{Q}}$ and $N^{\downarrow\downarrow}_{\nu,\boldsymbol{Q}}$ are individually in equilibrium and use that $\left|J^{\uparrow\uparrow}_{\boldsymbol{Q}}\right|^2 = \left|J^{\downarrow\downarrow}_{\boldsymbol{Q}}\right|^2$. Then, we define $\langle|J^s|^2\rangle_{s'} \equiv \langle|J|^2\rangle$ and obtain coupled differential equations for the spin-up and spin-down occupations

$$\begin{cases}\partial_\mathrm{t}\,\mathrm{N}^{\uparrow\uparrow} = -\dfrac{2}{\hbar\Gamma_{\text{x-ph}}}\langle|\mathrm{J}|^2\rangle\left(\mathrm{N}^{\uparrow\uparrow}-\mathrm{N}^{\downarrow\downarrow}\right) - \dfrac{\mathrm{N}^{\uparrow\uparrow}}{\tau_{\mathrm{dec}}}\\ \partial_\mathrm{t}\,\mathrm{N}^{\downarrow\downarrow} = -\dfrac{2}{\hbar\Gamma_{\text{x-ph}}}\langle|\mathrm{J}|^2\rangle\left(\mathrm{N}^{\downarrow\downarrow}-\mathrm{N}^{\uparrow\uparrow}\right) - \dfrac{\mathrm{N}^{\downarrow\downarrow}}{\tau_{\mathrm{dec}}}\end{cases} \tag{S14}$$

Introducing the total occupation $N^T = N^{\downarrow\downarrow} + N^{\uparrow\uparrow}$ as well as the difference in occupation $\delta N = N^{\downarrow\downarrow} - N^{\uparrow\uparrow}$ we obtain

$$\dot{\mathrm{N}}^\mathrm{T} = -\mathrm{N}^\mathrm{T}\tau_{\mathrm{dec}}\,,$$

$$\delta\dot{\mathrm{N}} = -\left(\frac{1}{\tau_{dec}} + \frac{1}{\tau_{spin}}\right)\delta\mathrm{N}\,, \tag{S15}$$

with the spin relaxation time

$$\tau_{\mathrm{spin}} = \frac{\hbar\Gamma_{\text{x-ph}}}{4\langle|\mathrm{J}|^2\rangle}. \tag{S16}$$

The equations for $\delta N$ and $N^T$ can be solved directly as $N^T(t) = N_0\,e^{-t/\tau_{dec}}$ and $\delta N(t) = N_0\,e^{-t/(1/\tau_{dec}+1/\tau_{spin})}$ assuming $N^T(0) \approx \delta N(0) = N_0$. The degree of circular polarization (DOP) is given by integrating the time-dependent occupations with respect to time resulting in

$$\text{DOP} = \frac{\int_0^\infty \mathrm{d}t\delta N(\mathrm{t})}{\int_0^\infty \mathrm{d}t N^{\mathrm{T}}(\mathrm{t})} = \frac{1}{1+\frac{\tau_{\mathrm{dec}}}{\tau_{\mathrm{spin}}}}. \tag{S17}$$

Importantly, as $\frac{\tau_{dec}}{\tau_{spin}} > 0$ it holds that DOP $\leq$ 1, and for $\tau_{spin} \to \infty$, the degree of circular polarization approaches unity. The results are formally similar to the ones obtained after continuous-wave excitation[17], with the intervalley exchange interaction now averaged over all the valleys, rather than just the bright one. Such a difference stems from the excitation scheme, which in the case of continuous-wave excitation constantly refills the bright KK states, which are hence the main source for the bright K′K′, whose dynamics in the static limit provides the DOP (namely $\partial_t \sum_Q N^{\downarrow\downarrow}_{\mathrm{K'K'},\boldsymbol{Q}} = 0$). We want to investigate the temperature dependence of the degree of circular polarization. By estimating the exciton occupation by a thermalized Boltzmann distribution, it follows

$$\langle |J|^2 \rangle = \frac{2J^2\,(M^{\mathrm{KK}})^2\,k_B T}{\hbar^2\,K^2 \sum_\nu \mathrm{e}^{\Delta E^\nu/(k_B T)}\,M^\nu\,g^\nu}, \tag{S18}$$

where $\Delta E^\nu = E^{\mathrm{KK}} - E^\nu$ is the energy difference between bright and (momentum)-dark states, $g^\nu$ is the valley degeneracy, $M^\nu$ is the total exciton mass. Furthermore, we used $J_Q = JQ/K$, where $J$ is the intervalley exchange coupling and $K$ =12.6 nm$^{-1}$, assuming an undoped system[18]. By noting that the long-range part of the momentum-dependent intervalley exchange interaction $J_Q = V_Q\,Q^2|M|^2$, where $M$ is the optical matrix element and assuming a simple Coulomb interaction of the form $V_Q = \frac{e^2}{2\epsilon_0\,\epsilon_s\,Q}$, $\epsilon_s$ being the background dielectric constant, we find that $J \approx 0.1$ eV with input parameters from Ref. [16]. We include explicitly the lowest-lying $\nu = \mathrm{KK}, \mathrm{KK'}$ and $\mathrm{K\Lambda}$ exciton states in h-BN-encapsulated 1L-$WSe_2$ and obtain the energy differences $\Delta E^{\mathrm{KK'}} = -46$ meV and $\Delta E^{\mathrm{K\Lambda}} = -34$ meV by solving the Wannier equation as detailed in Ref. [15]. The total exciton masses $M^\nu$ in each valley are extracted from Table VI in Ref. [16]. Furthermore, we note that the phonon-induced linewidth can be fitted according to

$$\Gamma_{x-ph} = c_1 T + c_2^{\mathrm{KK}} n_B(\Omega_{\mathrm{KK}}) + c_2^{\mathrm{KK'}}(n_B(\Omega_{\mathrm{KK'}}) + 1) + c_2^{\mathrm{K\Lambda}}(n_B(\Omega_{\mathrm{K\Lambda}}) + 1), \tag{S19}$$

aligning with the microscopic linewidth calculation of Ref. [19]. Here, the parameters read $c_1 = 13$ μeV/K, $c_2^{KK} = 3.5$ meV, $c_2^{KK'} = 1.4$ meV, $c_2^{K\Lambda} = 2.6$ meV, $\Omega_{KK} = 30.7$ meV with the intervalley phonon energies $\Omega'_{KK} = 16.8$ meV and $\Omega_{K\Lambda} = 12.95$ meV obtained after averaging the energies of the corresponding acoustic phonons inducing interband scattering. Furthermore, $n_B(\ldots)$ is the Bose-Einstein distribution. Finally, we find the temperature-dependent spin relaxation time:

$$\tau_{spin} = \frac{\hbar^3 K^2 \Gamma_{x-ph} \sum_\nu \mathrm{e}^{\Delta E^\nu/(k_B T)} M^\nu g^\nu}{8 J^2 (M^{KK})^2\, k_B T}. \tag{S20}$$

In general, the temperature dependence of the spin relaxation time is strongly influenced by the relative energetic alignment between bright and dark exciton states. If dark excitons are excluded and bright excitons are by far the energetically lowest states, we obtain a spin relaxation time $\tau_{spin} \propto \frac{1}{T}$, as found previously in literature[20]. In the case of $WSe_2$, the dark KK′ and KΛ states, however, constitute the energetically lowest states and the spin relaxation time becomes exponentially large at low temperatures (cf. exponential factors in Eq. S20) reflecting the suppression of the intervalley exchange scattering between momentum-dark states. With the temperature-dependent spin relaxation time at hand, the temperature-dependent degree of polarization is evaluated using Eq. S17 for a fixed temperature-independent decay time.

In Fig. S8, we present the degree of polarization for different decay times $\tau_{\mathrm{dec}}$ = 0.1, 0.2, 0.5, and 1 ns as a function of temperature. The temperature dependence of the DOP is found to be qualitatively similar for a broad range of decay times in agreement with previous calculations[16].

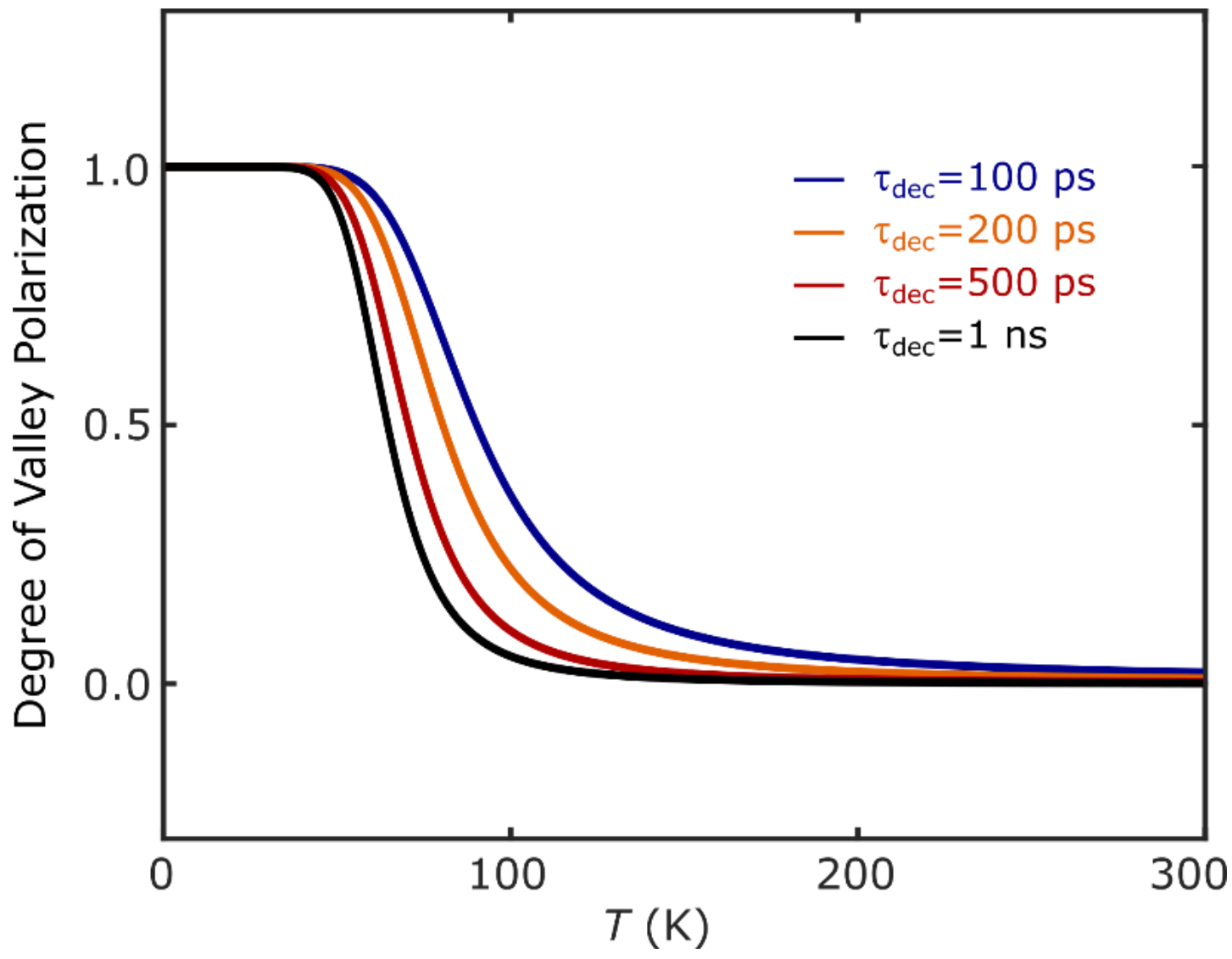

**Figure S8.** Degree of circular polarization as a function of temperature for different decay times, 0.1, 0.2, 0.5, and 1 ns.

## S11: Spatially-resolved photocurrent mappings.

To clarify the origin of the photocurrent in our 1L-$WSe_2$ devices, we perform scanning photocurrent mapping in device D1 (Fig S1f). This measurement enables us to identify where the photocurrent is generated in our device spatially. These spatially resolved photocurrent maps are measured at 50 K, with excitation wavelength of 690 nm, fluence of 30 μJ/cm$^2$, and $V_{ds}$ = 5 V. Figure S9 shows spatially resolved reflectivity (a and c) alongside their corresponding photocurrent maps (b and d). Figures S9a and S9b present the full-device reflectivity and photocurrent maps, respectively, providing an overview of how the photocurrent distribution relates to the device geometry. Figures S9c and S9d show higher resolution reflectivity and photocurrent maps of the same region, allowing a more precise visualization of the position of the generated photocurrent. The photocurrent mainly originates in the proximity of the metal contact/semiconductor interface, confirming that the in-built electric field due to the Schottky barriers allows for efficient dissociation of the generated excitons, increasing the photocurrent[21]. The presence of Schottky contacts in 1L-TMDs-based devices has been demonstrated to enable fast photoresponse times in the range of 1-100 ps[22].

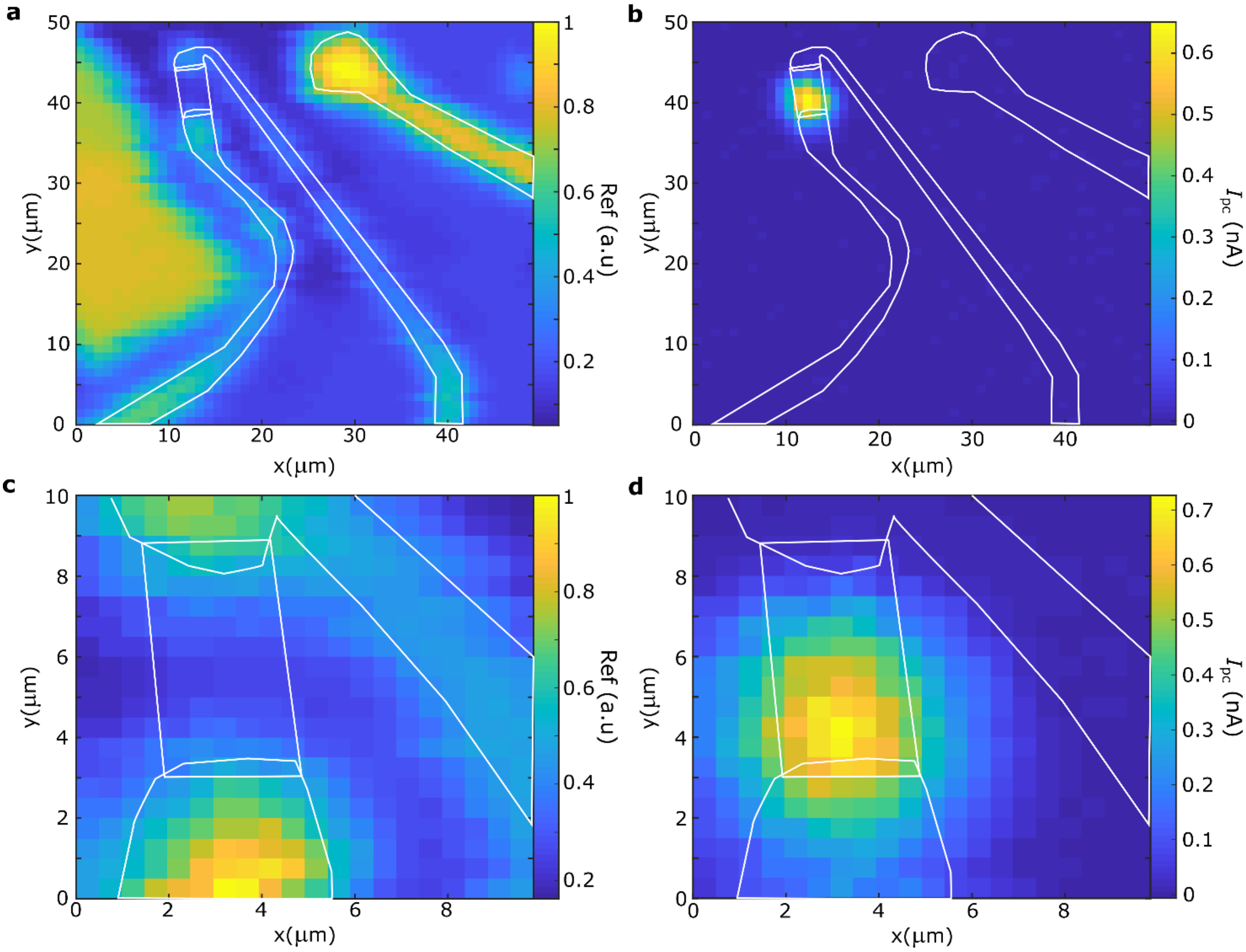

**Figure S9.** Spatially resolved photocurrent maps. (a) and (c) Spatially resolved normalized reflectivity with corresponding (b) and (c) spatially resolved photocurrent. Both measurements are taken simultaneously at a fixed drain-source voltage of 5 V and at an excitation wavelength of 690 nm.

Although the electrical contacts are not explicitly included in the theoretical modelling, Schottky contacts can affect the measured photocurrent in our devices, as seen in our spatially resolved photocurrent measurements. This in-plane electric field may therefore quantitatively modify both the photocurrent magnitude and its sublinearity. However, on previous studies of monolayer $WSe_2$ photodetectors, the sublinear response at the relevant exciton densities ($10^{11} < N_x < 10^{13}\mathrm{cm}^{-2}$) is still governed predominantly by exciton–exciton annihilation, even in the presence of strong in-plane electric fields[21]

In-plane electric fields are also expected to affect exciton decay radiative and nonradiative channels and thereby modify the valley degree of polarization. The decay time is captured phenomenologically in our model by the parameter $\tau_{dec}$. As shown in the previous Supporting Section, varying $\tau_{dec}$ over a broad range keeps unchanged the qualitative temperature dependence of the calculated degree of polarization. Specifically, calculations of the valley degree of polarization for different values of $\tau_{dec}$ give qualitatively similar behaviour (see Figure S8). These results indicate that the underlying physical mechanism governing the valley degree of polarization remains robust in the presence of Schottky contacts.